\documentclass[11pt]{article}

\usepackage[utf8]{inputenc}
\usepackage[T1]{fontenc}
\usepackage{amsmath,amssymb}
\usepackage{graphicx}
\usepackage{booktabs}
\usepackage{hyperref}
\usepackage{natbib}
\usepackage{xcolor}
\usepackage{enumitem}
\usepackage{microtype}
\usepackage{geometry}
\usepackage{tabularx}
\usepackage{multirow}
\usepackage{caption}
\usepackage{subcaption}
\usepackage{listings}

\hypersetup{
    colorlinks=true,
    linkcolor=blue,
    citecolor=blue,
    urlcolor=blue,
    pdftitle={SecureCode: A Production-Grade Multi-Turn Dataset for Training Security-Aware Code Generation Models},
    pdfauthor={Scott Thornton},
}

\title{\textbf{SecureCode: A Production-Grade Multi-Turn Dataset\\for Training Security-Aware Code Generation Models}}

\author{
Scott Thornton\\
\texttt{scott@perfecxion.ai}
}

\date{February 2026 (revised July 2026)}

\begin{document}

\maketitle

\begin{quote}
\small\noindent\textbf{Revision note (2026).} This version corrects and updates the February 2026 release of this paper. An independent audit found that the original ``100\% incident grounding'' claim was overstated and that the release splitting left measurable train/test leakage. In this version we re-verified every reference, relabeled representative scenarios, removed examples whose ``secure'' code did not fix the stated vulnerability, corrected the splits, and updated the dataset to its current audited form (2,372 examples: web 1,625, AI/ML 747). Section~\ref{sec:audit} documents the audit. Where a figure still depicts the February 2026 composition, the corrected totals in Section~\ref{sec:audit} and the dataset cards on HuggingFace are authoritative.
\end{quote}

% ===== Abstract =====
\begin{abstract}
AI coding assistants produce vulnerable code in 45\% of security-relevant scenarios \citep{veracode2025}, yet no public training dataset teaches both traditional web security and AI/ML-specific defenses in a format suitable for instruction tuning. We present \textbf{SecureCode}, a production-grade dataset of \textbf{2,372 multi-turn security training examples} spanning two domains: web application security (1,625 examples covering the OWASP Top 10 2021 across 11 languages and 9 frameworks) and AI/ML security (747 examples covering all 10 OWASP LLM Top 10 2025 categories across 40+ frameworks including LangChain, OpenAI, and HuggingFace). A 2026 grounding audit (Section~\ref{sec:audit}) independently re-verified every reference and corrected the dataset's grounding claims: we report the fraction of examples tied to a documented incident or verified CVE rather than asserting complete grounding. Every example follows a 4-turn conversational structure---feature request, vulnerable and secure implementations with attack demonstrations, advanced probing, and defense-in-depth operational guidance---designed for direct use in instruction tuning pipelines.

Quality assurance combines automated structural validation with multi-agent review from 7 specialist AI perspectives (10,500+ assessments) and an 8-phase remediation pipeline, producing a rubric-calibrated mean quality score of 93.8/100 ($\sigma = 0.93$) for the AI/ML component. Each example provides SIEM integration strategies, infrastructure hardening recommendations, and testing approaches using production frameworks.

We release the unified dataset on HuggingFace with domain-specific loading configurations (\texttt{web}, \texttt{aiml}, \texttt{default}), alongside 8 fine-tuned open-source models (3B--20B parameters, QLoRA), and an evaluation framework with four security-specific metrics. To our knowledge, SecureCode is the first public dataset that jointly provides OWASP Top 10 2021 web coverage and OWASP LLM Top 10 2025 AI/ML coverage in a unified conversational schema suitable for instruction tuning.
\end{abstract}

% ===== 1. Introduction =====
\section{Introduction}

\subsection{The Security Crisis in AI-Generated Code}

AI coding assistants produce vulnerable code in 45\% of generated implementations \citep{veracode2025}. Veracode's 2025 GenAI Code Security Report analyzed code from leading AI assistants and found nearly half of security-relevant implementations contained Common Weakness Enumeration (CWE) vulnerabilities. This represents systematic risk in AI-assisted development, compounding security debt across millions of developers.

The issue goes beyond individual bugs. AI-generated vulnerabilities enter production codebases silently, without the traditional code review scrutiny applied to human-written code. Developers trust AI assistants to produce functional implementations, but these tools lack the security context needed to recognize when ``functional'' means ``exploitable.'' Apiiro (2025) found AI copilots introduced 322\% more privilege escalation paths and 153\% more architectural design flaws compared to manually-written code, while generating 10$\times$ more security findings overall---actively degrading security practices \citep{apiiro2025}.

This creates a multiplier effect. Vulnerable patterns suggested by AI assistants propagate across multiple projects. SQL injection flaws spread through microservices architectures. Authentication bypasses replicate across API endpoints. Cryptographic failures multiply through mobile applications. The scale of AI adoption makes this a systematic risk to software security.

LLMs trained on public code repositories learn from millions of vulnerable examples. Stack Overflow answers from 2010 showing insecure MySQL queries. GitHub repositories implementing broken authentication. Tutorial code demonstrating SQL injection vulnerabilities as ``simple examples.'' These models learn what code \emph{looks like}, not what secure code \emph{requires}.

\subsection{The Emerging AI/ML Security Gap}

The security crisis extends beyond traditional web vulnerabilities. Organizations deploy Large Language Models at unprecedented speed---the global AI market reached \$184 billion in 2024 \citep{grandview2024}---yet the AI coding assistants building these systems have a critical blind spot. They were trained on millions of code examples, but almost none of that training data addresses AI-specific security vulnerabilities.

When a developer asks ``how do I build a RAG pipeline with LangChain?'', the model produces functional code without input validation, embedding sanitization, or output filtering. The model was never shown what can go wrong. Prompt injection attacks increased over 300\% between 2024 and 2025 \citep{owasp2025llm}. RAG poisoning emerged as a practical attack vector against enterprise knowledge systems. In one documented case, criminals used deepfake technology to steal \$25.6 million from a single company by impersonating executives in a video conference \citep{arup2024}.

The OWASP Foundation recognized the scale of this problem by publishing the \textbf{OWASP LLM Top 10 2025}, identifying 10 critical vulnerability categories specific to LLM applications \citep{owasp2025llm}. These categories---from prompt injection (LLM01) to unbounded consumption (LLM10)---define an entirely new attack surface that no existing training dataset addresses.

\subsection{Why Existing Datasets Fall Short}

Existing secure coding datasets have significant limitations for training security-aware language models. We analyzed four widely-used datasets: CWE-Sans (372 examples), Juliet Test Suite ($\sim$81,000--86,000 synthetic test cases for C/C++ and Java), SARD ($\sim$170,000--200,000 test programs), and Draper VDISC (1.27 million C examples). While these serve their intended purposes, they have critical gaps for LLM training.

\textbf{Scale versus quality.} Juliet provides $\sim$81,000--86,000 test cases designed for testing static analysis tools, but lacks connections to real-world incidents \citep{juliet2012}. SARD offers $\sim$170,000--200,000 test programs but fewer than 5\% reference documented security incidents \citep{sard}. Synthetic training data misses the contextual factors that make vulnerabilities exploitable in production.

\textbf{Incident grounding is rare.} Our manual audit of CWE-Sans metadata (n=372 examples, 100\% coverage) found approximately 18\% reference actual CVEs or documented breaches---fewer than one in five \citep{cwe2019}. Real-world attacks exploit edge cases, framework-specific behaviors, and integration failures that rarely appear in manufactured examples.

\textbf{Format limitations.} Existing datasets use code-only formats---vulnerable snippet paired with secure snippet. This doesn't capture how developers actually interact with AI assistants. Real development conversations escalate through multiple turns as developers ask about functionality, scaling, performance, and edge cases. AI assistants must maintain security context throughout this workflow, but existing datasets don't model these multi-turn interactions.

\textbf{No operational guidance.} Existing datasets provide vulnerable and patched code without detection mechanisms, logging strategies, or defense-in-depth guidance. For production systems, secure code is just one component of comprehensive security.

\textbf{No AI/ML coverage.} Most critically, no existing public dataset addresses AI-specific vulnerability classes. When an attacker poisons a RAG database to make an LLM recommend transferring funds to a fraudulent account, that is not SQL injection. When a prompt injection extracts a system prompt containing proprietary business logic, that is not XSS. AI/ML introduces fundamentally new vulnerability classes---prompt injection, data poisoning, model extraction, excessive agency, vector embedding attacks---that require purpose-built training data.

\subsection{SecureCode: A Unified Solution}

We developed SecureCode to address these limitations with a unified, production-grade training dataset spanning both traditional and AI/ML security domains. The dataset provides \textbf{2,372 rigorously validated examples} (post-audit; Section~\ref{sec:audit}) organized into two complementary components:

\textbf{Web Security (1,625 examples).} Covering the OWASP Top 10 2021 across 11 programming languages (Python, JavaScript, Java, Go, PHP, C\#, TypeScript, Ruby, Rust, Kotlin, YAML) and 9 web frameworks. The 1,216-example baseline achieved 100\% \emph{structural} compliance through systematic validation, and 219 framework-specific examples expanded coverage to production stacks (Express/NestJS, Laravel/Symfony, Spring Boot, Gin, Rails, ASP.NET Core). A subsequent 2026 audit (Section~\ref{sec:audit}) re-verified references, removed defective examples, and added current coverage, bringing the component to 1,625; grounding is reported explicitly rather than asserted for every example.

\textbf{AI/ML Security (747 examples).} Covering all 10 OWASP LLM Top 10 2025 categories (75 examples each; 74 in three categories after the 2026 fix-correctness audit removed 3 examples whose secure code did not mitigate the stated vulnerability), spanning 40+ AI/ML frameworks (LangChain, OpenAI, Anthropic, HuggingFace, LlamaIndex, FastAPI, Flask, ChromaDB, Pinecone, and many more) across 8 programming languages. Quality assurance combined multi-agent review from 7 specialist perspectives (10,500+ assessments) with an 8-phase remediation pipeline, achieving a mean quality score of 93.8/100.

\begin{figure}[t]
\centering
\includegraphics[width=\textwidth]{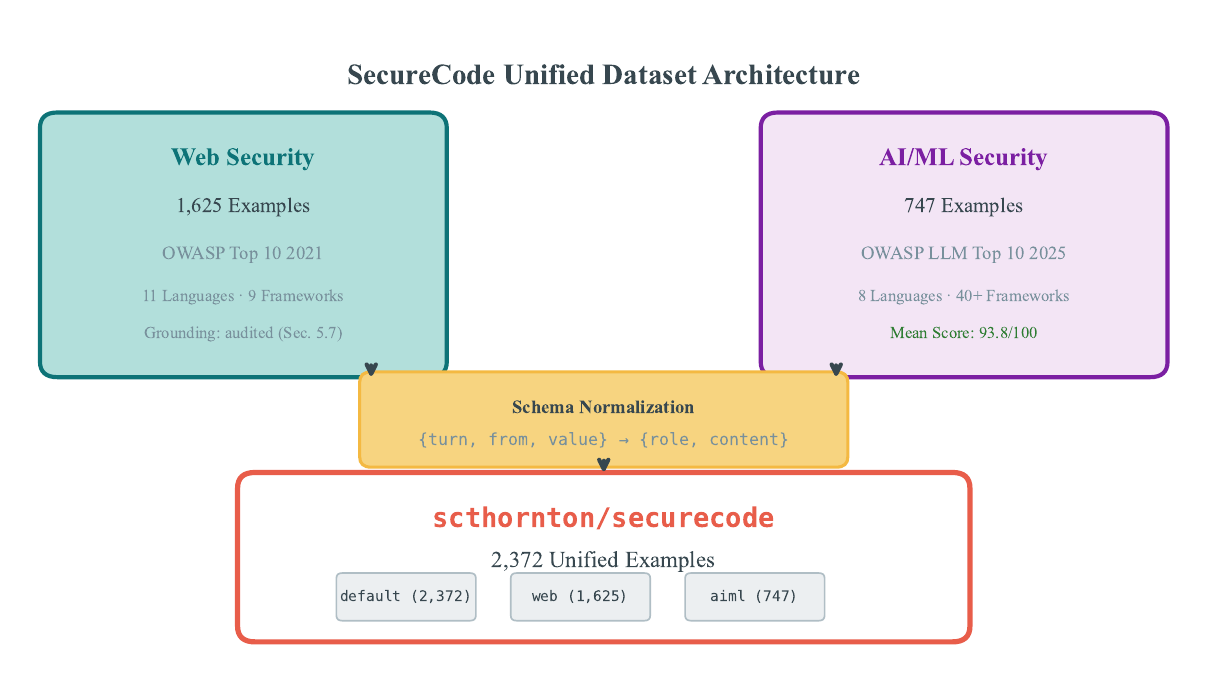}
\caption{SecureCode Unified Dataset Architecture. The web security component (1,625 examples, OWASP Top 10 2021) and AI/ML security component (747 examples, OWASP LLM Top 10 2025) are unified under a common schema with flexible HuggingFace loading configurations.}
\label{fig:architecture}
\end{figure}

Both components share a 4-turn conversational structure mirroring actual developer--AI interactions, providing vulnerable and secure implementations alongside attack demonstrations and defense-in-depth operational guidance. The unified dataset is available on HuggingFace (\texttt{scthornton/securecode}) with three loading configurations: \texttt{default} (all 2,372 examples), \texttt{web} (1,625 web security), and \texttt{aiml} (747 AI/ML security). Figure~\ref{fig:architecture} illustrates this unified architecture.

\subsection{Contributions}

This paper makes seven contributions to secure AI-assisted development:

\begin{enumerate}[nosep]
    \item \textbf{Unified security training dataset (2,372 examples).} To our knowledge, the first public dataset that jointly provides (i) OWASP Top 10 2021 web-application coverage and (ii) OWASP LLM Top 10 2025 AI/ML coverage, (iii) in a unified conversational schema suitable for instruction tuning---covering 20 vulnerability categories across 12+ programming languages and 40+ frameworks.

    \item \textbf{8 fine-tuned open-source models.} A collection of security-specialized models ranging from 3B to 20B parameters, trained on the unified dataset using QLoRA 4-bit quantization, released on HuggingFace.

    \item \textbf{Multi-agent quality assurance pipeline.} 7 specialist AI reviewers producing 10,500+ assessments combined with an 8-phase remediation pipeline, achieving a mean quality score of 93.8/100 ($\sigma = 0.93$) for the AI/ML component.

    \item \textbf{OWASP taxonomy alignment methodology.} A systematic approach to correcting category mismatches through staging-based file rotation, targeted generation, and overflow archival.

    \item \textbf{Schema unification framework.} A normalization pipeline converting heterogeneous conversation formats into a unified \texttt{\{role, content\}} schema with consistent metadata, enabling cross-domain training.

    \item \textbf{Novel 4-turn conversational structure.} A conversation format (feature request $\rightarrow$ vulnerable/secure implementations $\rightarrow$ advanced scenarios $\rightarrow$ operational guidance) that trains models on realistic developer--AI workflows.

    \item \textbf{Open-source release.} Dataset, component datasets, validation tools, training scripts, and 8 fine-tuned models released under CC~BY-NC-SA~4.0 (dataset) and open-source licenses (models) on HuggingFace and GitHub.
\end{enumerate}

\textbf{Claim boundary.} By ``unified'' we mean a single HuggingFace dataset with a common \texttt{\{role, content\}} conversation schema and shared loading interface across both security domains. By ``AI/ML security training examples'' we mean multi-turn code demonstrations---vulnerable implementation, secure alternative, attack explanation, and operational guidance---grounded in documented attack research. We exclude general prompt-injection text datasets that lack code, pure vulnerability corpora without conversational structure, and toxicity/content-moderation datasets that address text safety rather than code security.

% ===== 2. Related Work =====
\section{Related Work}

\subsection{Secure Coding Datasets}

The security research community has produced several datasets for studying vulnerable code, but none meet the combined requirements for training production-grade AI coding assistants across both traditional and AI/ML security domains.

\textbf{CWE-Sans Top 25 Dataset} provides 372 examples across 4 programming languages with partial OWASP coverage \citep{cwe2019}. Only 18\% of examples anchor to real-world incidents---the remaining 82\% are synthetic demonstrations of CWE patterns. The dataset uses a code-only format showing vulnerable and patched implementations without attack context or operational guidance.

\textbf{Juliet Test Suite} offers $\sim$81,000--86,000 synthetic test cases in C/C++ and Java covering 118 CWE types \citep{juliet2012}. Zero percent ground to real-world incidents. Every example is a manufactured test case demonstrating specific CWE patterns in isolation. The suite serves its intended purpose---testing static analysis tools---but synthetic examples miss the framework-specific quirks, integration failures, and configuration mistakes that cause actual breaches.

\textbf{Software Assurance Reference Dataset (SARD)} contains $\sim$170,000--200,000 test programs across 5 languages with no OWASP mapping \citep{sard}. Fewer than 5\% tie to documented security incidents. SARD focuses on test cases for automated analysis tools, not training data for AI models.

\textbf{Draper VDISC} provides 1.27 million C examples for binary analysis research \citep{vdisc}. This massive dataset concentrates entirely on C without multi-language coverage and lacks the high-level security context needed for training AI coding assistants.

\begin{figure}[t]
\centering
\includegraphics[width=\textwidth]{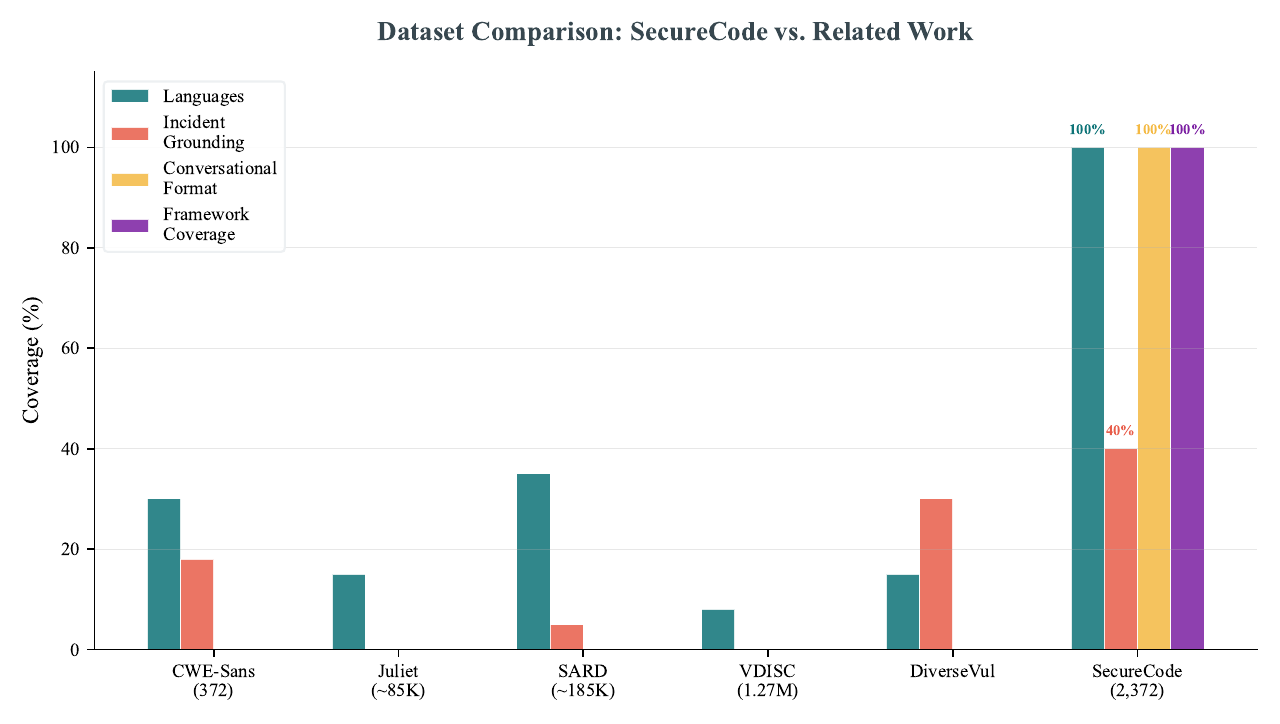}
\caption{Dataset Comparison. SecureCode vs.\ related work across four dimensions: language coverage, incident grounding, conversational format, and framework coverage. SecureCode is the only dataset combining a multi-turn conversational format with comprehensive framework coverage; its incident grounding is partial and independently audited (Section~\ref{sec:audit}) rather than assumed.}
\label{fig:comparison}
\end{figure}

Figure~\ref{fig:comparison} summarizes this comparison across four dimensions. \textbf{DiverseVul and BigVul} focus on C/C++ memory safety vulnerabilities---buffer overflows, use-after-free, integer overflows. While valuable for systems programming security, they do not cover the application-layer or AI/ML vulnerabilities that dominate modern deployments.

\subsection{AI Code Generation Security Research}

Recent empirical studies confirm that AI coding assistants systematically produce insecure code.

Veracode (2025) evaluated leading AI coding assistants and found 45\% of AI-generated implementations in security-relevant contexts contained CWE vulnerabilities \citep{veracode2025}. SQL injection appeared in database query generations. Command injection emerged in system interaction code. Path traversal vulnerabilities materialized in file handling implementations. Yet no training dataset existed to retrain these models on correct implementations.

Apiiro (2025) analyzed application security across thousands of repositories in Fortune 50 enterprises \citep{apiiro2025}. AI copilots introduced 322\% more privilege escalation paths and 153\% more architectural design flaws, generating 10$\times$ more security findings overall compared to manually-written code.

\subsection{AI/ML Security Research}

The AI security research community has documented a growing attack surface specific to LLM applications.

\textbf{Prompt injection} was formalized by Perez and Ribeiro (2022) and has since evolved from simple direct injections to sophisticated indirect attacks through RAG documents, tool descriptions, and multi-modal inputs \citep{greshake2023,perez2022}. The OWASP LLM Top 10 2025 classifies prompt injection as the highest-risk vulnerability category.

\textbf{RAG poisoning} attacks manipulate vector database contents to influence LLM outputs. Zou et al.\ (2023) demonstrated that adversarial suffixes transfer across models, while Chaudhari et al.\ (2024) showed embedding space manipulation can surface attacker-controlled content in retrieval results \citep{zou2023}.

\textbf{Model extraction and inversion} attacks probe deployed models through APIs to reconstruct model weights or training data. Carlini et al.\ (2021) showed that GPT-2 memorizes and regurgitates training data, raising privacy concerns for fine-tuned models handling sensitive information \citep{carlini2021}.

\textbf{Supply chain attacks} target the AI/ML dependency ecosystem---compromised model weights on HuggingFace, typosquatted packages on PyPI, and poisoned training data pipelines. \textbf{Agentic AI risks} emerge as LLM applications gain autonomous capabilities through over-permissioned tool use, unauthorized actions, and self-replicating agent patterns.

Despite this rich research landscape, no dataset previously translated these findings into training data for AI coding assistants. SecureCode bridges this gap.

\subsection{OWASP Standards}

SecureCode maps to two complementary OWASP taxonomies:

\textbf{OWASP Top 10 2021} defines the 10 most critical web application security risks, from broken access control (A01) to server-side request forgery (A10). The web security component of SecureCode provides complete coverage of these categories.

\textbf{OWASP LLM Top 10 2025} defines 10 critical vulnerability categories specific to LLM applications \citep{owasp2025llm}: prompt injection (LLM01), sensitive information disclosure (LLM02), supply chain vulnerabilities (LLM03), data and model poisoning (LLM04), improper output handling (LLM05), excessive agency (LLM06), system prompt leakage (LLM07), vector and embedding weaknesses (LLM08), misinformation (LLM09), and unbounded consumption (LLM10). The AI/ML component provides complete coverage with 75 examples per category.

% ===== 3. Dataset Design =====
\section{Dataset Design}

\subsection{Design Principles}

Five principles guided the design of SecureCode across both security domains:

\textbf{P1: Grounding, verified.} Examples reference documented security incidents, CVEs, vendor advisories, or published attack research where such a source exists; where it does not, the example is labeled a representative scenario rather than presented as a documented incident. Grounding is independently verified rather than assumed (Section~\ref{sec:audit}). Rather than manufacturing hypothetical vulnerabilities, we study actual breaches and documented attack techniques, extract vulnerable patterns, and build examples demonstrating both the vulnerability and the secure alternative.

\textbf{P2: Conversational Structure.} Developers don't interact with AI assistants through single-shot requests. They iterate---asking for basic functionality, evaluating the response, then probing deeper into scaling, performance, edge cases, and security hardening. Our 4-turn conversational structure captures this iterative workflow.

\textbf{P3: Dual Implementation Pattern.} Every example provides vulnerable and secure implementations of the same functionality. This side-by-side comparison enables contrastive learning---models learn what makes code insecure by seeing the exact pattern to avoid, then immediately learn the secure alternative with 5+ layered defenses.

\textbf{P4: Operational Completeness.} Security doesn't end at secure code. Production systems need detection, monitoring, incident response, and graceful degradation. Every example includes SIEM integration strategies, logging recommendations, and deployment hardening guidance.

\textbf{P5: Framework Authenticity.} Code uses real framework APIs with correct import paths, method signatures, and configuration patterns. A LangChain example uses \texttt{from langchain.chains import RetrievalQA} with proper chain configuration. A Spring Boot example uses \texttt{@RestController} with actual dependency injection. No pseudocode or simplified abstractions.

\subsection{Four-Turn Conversation Structure}

\begin{figure}[t]
\centering
\includegraphics[width=0.9\textwidth]{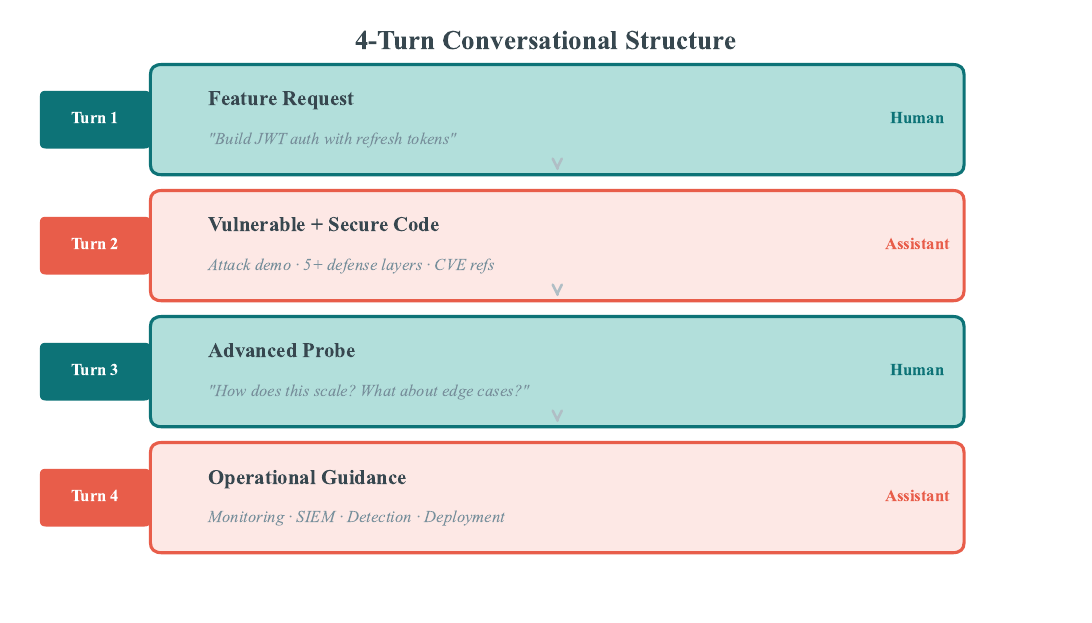}
\caption{Four-Turn Conversational Structure. Each example follows this pattern across both web and AI/ML security domains: developer question, vulnerable + secure implementations with attack demonstrations, follow-up probe, and production-grade operational guidance.}
\label{fig:conversation}
\end{figure}

Every example in the unified dataset follows a 4-turn conversation that teaches security through natural developer interaction (Figure~\ref{fig:conversation}):

\textbf{Turn 1 (Human):} A developer asks how to build a specific feature. The question is natural and specific---the kind of question developers paste into AI coding assistants. \emph{``How do I implement JWT authentication with refresh tokens?''} or \emph{``How do I build a RAG pipeline with LangChain and Pinecone?''}

\textbf{Turn 2 (Assistant):} The response shows a vulnerable implementation first (clearly marked), explains the specific risks and how an attacker would exploit them, then provides a secure implementation with multiple defense layers. Real-world incidents and CVE references ground the security claims.

\textbf{Turn 3 (Human):} A follow-up probing deeper---testing strategies, edge cases, advanced attack scenarios, scaling concerns, or deployment questions. \emph{``How does this scale to 10,000 concurrent users?''} or \emph{``How would I test this for indirect prompt injection in production?''}

\textbf{Turn 4 (Assistant):} Production-grade guidance covering operational security: logging, monitoring, detection rules, SIEM integration, infrastructure hardening, deployment considerations, and common mistakes to avoid.

The following excerpt illustrates a condensed AI/ML security example (LLM01: Prompt Injection):

\begin{quote}
\small
\textbf{Turn 1 (Human):} ``How do I build a RAG pipeline with LangChain and Pinecone that lets users query our internal documentation?''

\textbf{Turn 2 (Assistant):} \emph{[Vulnerable implementation:} unvalidated user input passed directly to \texttt{RetrievalQA.from\_chain\_type()}\emph{]} $\rightarrow$ \emph{[Secure implementation:} input sanitization, embedding similarity threshold, output filtering, prompt armoring, content boundary tags\emph{]} $\rightarrow$ \emph{[Attack:} indirect injection via poisoned document embedding\emph{]}

\textbf{Turn 3 (Human):} ``How would I test this for indirect prompt injection, and what monitoring should I set up in production?''

\textbf{Turn 4 (Assistant):} \emph{[Testing:} adversarial document injection test suite, boundary escape fuzzing\emph{]} $\rightarrow$ \emph{[Monitoring:} semantic drift detection, output consistency scoring, anomaly alerting with PagerDuty/Slack integration\emph{]}
\end{quote}

\subsection{Web Security Domain}

\begin{figure}[h]
\centering
\includegraphics[width=\textwidth]{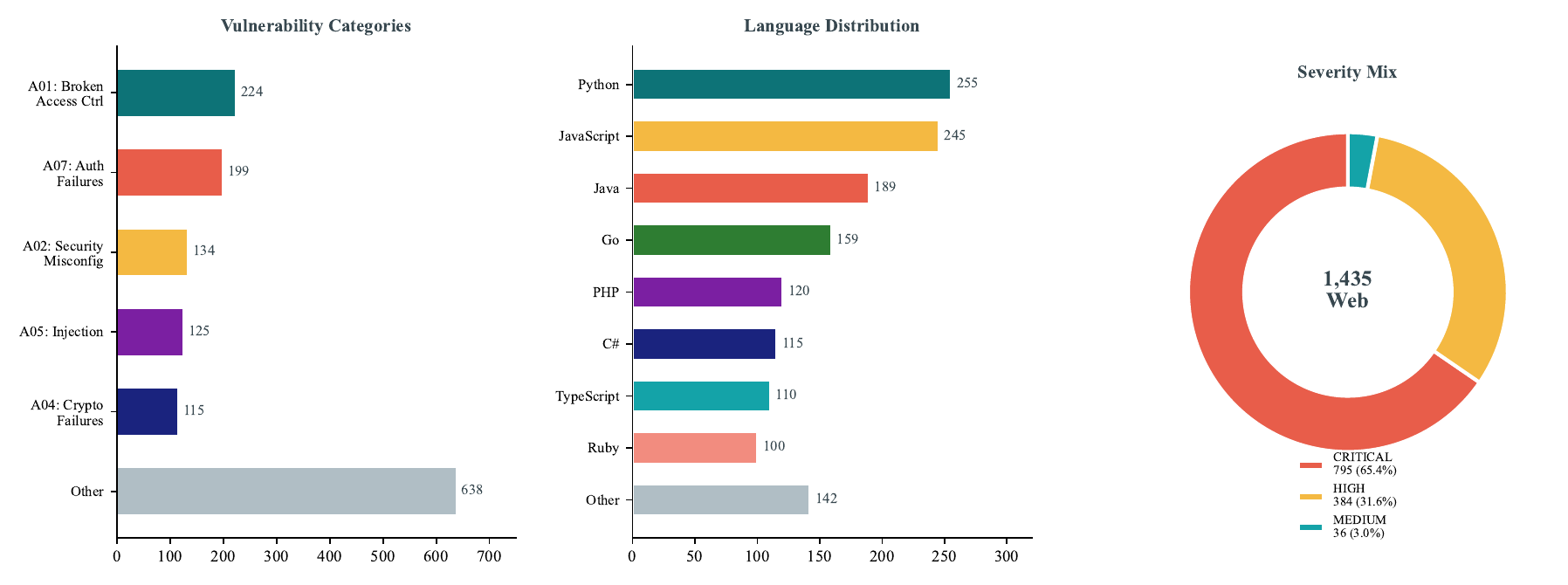}
\caption{Web Security Coverage Snapshot. Dataset composition across three dimensions: vulnerability categories (left), language distribution (center), and severity mix (right). (February 2026 composition; current per-category counts appear in Table~\ref{tab:web-coverage}.)}
\label{fig:coverage}
\end{figure}

The web security component provides 1,625 examples covering the OWASP Top 10 2021 (Table~\ref{tab:web-coverage}, Figure~\ref{fig:coverage}), organized into two tiers:

\textbf{Baseline (1,216 examples).} Covers 11 vulnerability categories across 11 programming languages (Python, JavaScript, Java, Go, PHP, C\#, TypeScript, Ruby, Rust, Kotlin, YAML for infrastructure-as-code). Examples reference documented CVEs or security incidents---from the 2017 Equifax breach (CVE-2017-5638) costing \$425 million to the 2019 Capital One SSRF attack exposing 100 million customer records---though a 2026 audit (Section~\ref{sec:audit}) found many of these references required correction. Structural compliance progressed from 47.2\% baseline to 100\% through systematic remediation across five fix categories.

\textbf{Framework Expansion (219 examples).} Adds framework-specific implementations ensuring authentic production patterns: Express/NestJS for JavaScript, Laravel/Symfony for PHP, Spring Boot for Java, Gin for Go, Rails for Ruby, and ASP.NET Core for C\#. These examples demonstrate the same vulnerability classes through framework-native idioms and APIs.

\begin{table}[h]
\centering
\caption{Web Security Coverage (OWASP Top 10 2021)}
\label{tab:web-coverage}
\begin{tabular}{@{}lrr@{}}
\toprule
\textbf{Category} & \textbf{Examples} & \textbf{Percentage} \\
\midrule
A03: Injection & 303 & 18.6\% \\
A01: Broken Access Control & 303 & 18.6\% \\
A07: Identification and Authentication Failures & 234 & 14.4\% \\
A05: Security Misconfiguration & 158 & 9.7\% \\
A04: Insecure Design & 138 & 8.5\% \\
A06: Vulnerable and Outdated Components & 137 & 8.4\% \\
A02: Cryptographic Failures & 135 & 8.3\% \\
A08: Software and Data Integrity Failures & 99 & 6.1\% \\
A09: Security Logging and Monitoring Failures & 66 & 4.1\% \\
A10: Server-Side Request Forgery (SSRF) & 52 & 3.2\% \\
\midrule
\textbf{Total (Web)} & \textbf{1,625} & \textbf{100\%} \\
\bottomrule
\end{tabular}
\end{table}

\subsection{AI/ML Security Domain}

The AI/ML security component provides 747 examples covering all 10 OWASP LLM Top 10 2025 categories, 75 per category except three reduced to 74 by the fix-correctness audit (Table~\ref{tab:aiml-categories}; Section~\ref{sec:audit}).

\begin{table}[h]
\centering
\caption{AI/ML Security Coverage (OWASP LLM Top 10 2025)}
\label{tab:aiml-categories}
\begin{tabular}{@{}llrr@{}}
\toprule
\textbf{Code} & \textbf{Category} & \textbf{Examples} & \textbf{Mean Score} \\
\midrule
LLM01 & Prompt Injection & 75 & 94.0 \\
LLM02 & Sensitive Information Disclosure & 74 & 93.8 \\
LLM03 & Supply Chain Vulnerabilities & 75 & 93.9 \\
LLM04 & Data and Model Poisoning & 75 & 93.8 \\
LLM05 & Improper Output Handling & 75 & 93.2 \\
LLM06 & Excessive Agency & 74 & 93.8 \\
LLM07 & System Prompt Leakage & 74 & 93.6 \\
LLM08 & Vector and Embedding Weaknesses & 75 & 94.0 \\
LLM09 & Misinformation & 75 & 93.8 \\
LLM10 & Unbounded Consumption & 75 & 93.5 \\
\midrule
\textbf{Total (AI/ML)} & & \textbf{747} & \textbf{93.8} \\
\bottomrule
\end{tabular}
\end{table}

\begin{figure}[h]
\centering
\includegraphics[width=\textwidth]{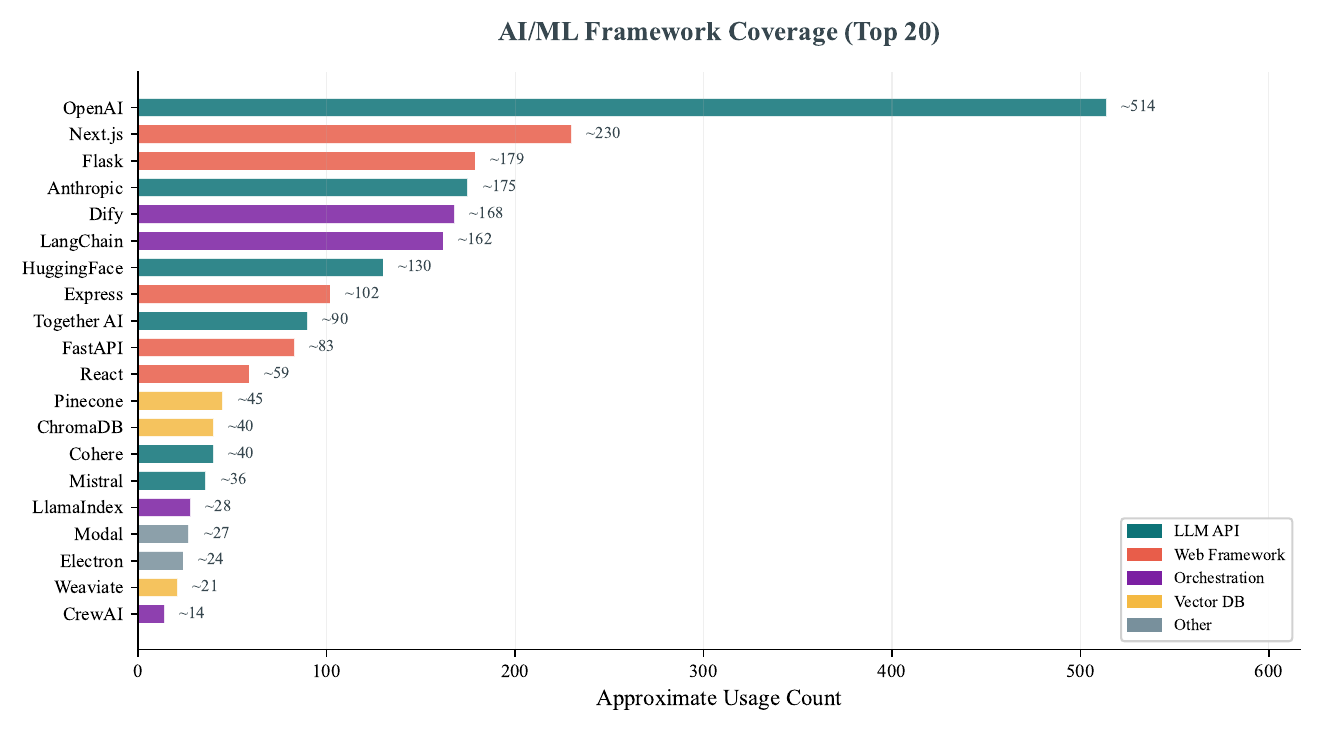}
\caption{AI/ML Framework Coverage (Top 20). Framework usage colored by category: LLM APIs (teal), web frameworks (coral), orchestration (purple), and vector databases (gold). Most examples use 2--4 frameworks together.}
\label{fig:frameworks}
\end{figure}

The AI/ML component covers 40+ frameworks organized by function (Figure~\ref{fig:frameworks}): LLM APIs (OpenAI, Anthropic, HuggingFace, Mistral, Cohere, Together AI, Groq, DeepSeek), orchestration frameworks (LangChain, LlamaIndex, CrewAI, AutoGen, Dify), vector databases (ChromaDB, Pinecone, Qdrant, Weaviate, Milvus, FAISS), web frameworks (FastAPI, Flask, Django, Express, Next.js), serving platforms (vLLM, BentoML, Ray Serve, Modal, MLflow), and UI frameworks (Gradio, Streamlit, Chainlit). Python dominates at 90.7\% (680 examples), reflecting the reality of AI/ML development, with TypeScript (4.7\%) and JavaScript (4.0\%) covering frontend and API layers.

\subsection{Schema Unification}
\label{sec:unification}

The web security component (originally SecureCode v2) used a \texttt{\{turn, from, value\}} conversation schema, while the AI/ML component used \texttt{\{role, content\}}. To create a unified training dataset, we normalized all examples to the \texttt{\{role, content\}} format used by modern LLM training pipelines:

\begin{lstlisting}[basicstyle=\ttfamily\footnotesize]
// Web security (original)         // Unified format
{"turn": 1,                   -->  {"role": "human",
 "from": "human",                   "content": "..."}
 "value": "..."}
\end{lstlisting}

The unification pipeline preserves all domain-specific metadata (OWASP categories, CWE mappings, severity levels, quality scores, references) while ensuring a consistent top-level conversation format. The resulting dataset is available on HuggingFace with three loading configurations:

\begin{itemize}[nosep]
    \item \texttt{default} --- All 2,372 examples (web + AI/ML)
    \item \texttt{web} --- 1,625 web security examples only
    \item \texttt{aiml} --- 747 AI/ML security examples only
\end{itemize}

This flexibility allows researchers to train on the full dataset, focus on a single domain, or compare model performance across security domains.

% ===== 4. Dataset Construction =====
\section{Dataset Construction}

\subsection{Web Security Examples}

Web security examples were constructed through a three-phase methodology ensuring incident grounding and production quality.

\textbf{Phase 1: Incident mining.} We mined CVE databases from 2017--2025, analyzed OWASP Top 10 documentation, reviewed security breach reports, and studied bug bounty disclosures. Each example ties to a specific incident: the 2017 Equifax breach (CVE-2017-5638) from Apache Struts 2 RCE, the 2019 Capital One SSRF attack exposing 100 million records, and deserialization vulnerabilities that compromised dozens of financial institutions.

\textbf{Phase 2: Example generation.} Scenarios were generated using multi-LLM synthesis (ChatGPT, Claude, Llama) with human expert review. Every example provides vulnerable and secure implementations with attack demonstrations and operational guidance. The code implementations are synthetically generated; the incidents are real.

\textbf{Phase 3: Validation and expansion.} An automated validation framework (\texttt{validate\_contributing\_compliance.py}) enforced structural quality: 4-turn conversation compliance, CVE formatting, language tag validity, and content quality. Compliance progressed from 47.2\% to 100\% through systematic remediation of 604 specific issues. A subsequent framework expansion phase added 219 examples covering production web frameworks (Express/NestJS, Laravel/Symfony, Spring Boot, Gin, Rails, ASP.NET Core), bringing the total from 1,216 to 1,435.

\begin{figure}[t]
\centering
\includegraphics[width=\textwidth]{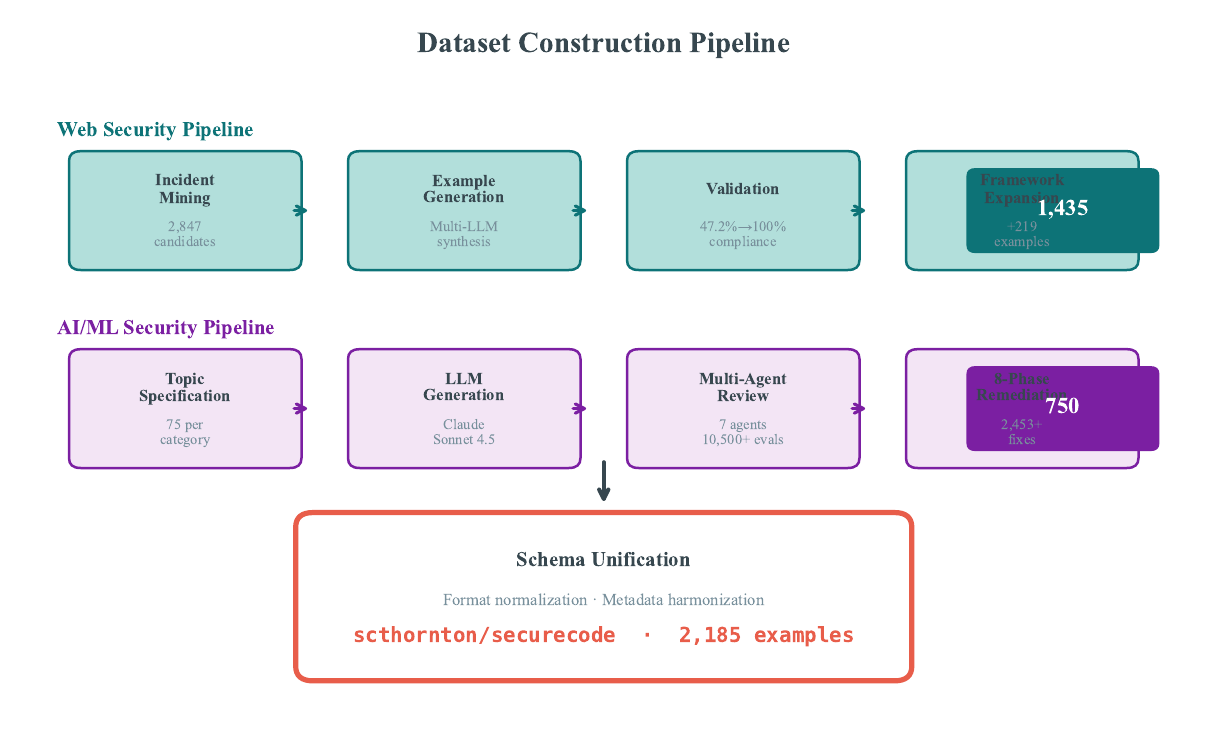}
\caption{Dataset Construction Pipeline. The web security component progresses from incident mining through generation and validation. The AI/ML component follows a parallel pipeline with multi-agent review and 8-phase remediation. Both feed into the unified dataset through schema normalization. (Construction as of February 2026; the dataset was subsequently audited to 2,372 examples, Section~\ref{sec:audit}.)}
\label{fig:pipeline}
\end{figure}

\subsection{AI/ML Security Examples}

AI/ML security examples were constructed through a multi-stage pipeline combining LLM-based generation with systematic quality assurance.

\textbf{Stage 1: Topic specification.} For each OWASP LLM Top 10 2025 category, we defined 75 unique vulnerability scenarios specifying the target framework, attack technique, subcategory, and expected defense layers. Topics were selected to maximize technique diversity (704 unique techniques across 750 files) and framework coverage (40+ frameworks).

\textbf{Stage 2: Example generation.} Each scenario was generated using Claude Sonnet 4.5 with a structured prompt enforcing the 4-turn conversational format, dual implementation pattern, 5+ defense layers, and grounded references.

\textbf{Stage 3: Immediate validation.} Every generated file was validated immediately: JSON parse check, schema compliance, 4-turn structure, and quality score threshold. Files failing validation were regenerated with adjusted prompts.

\textbf{Stage 4: Batch quality assurance.} All 750 files underwent multi-agent review and 8-phase remediation (Sections~\ref{sec:multi-agent}--\ref{sec:remediation}). Figure~\ref{fig:pipeline} shows the complete dual construction pipeline.

\subsection{Multi-Agent Quality Review}
\label{sec:multi-agent}

Every AI/ML security example was reviewed by 7 specialist AI agents, each evaluating from a distinct perspective (Table~\ref{tab:agents}).

\begin{table}[h]
\centering
\caption{Multi-Agent Review: 7 Specialist Perspectives}
\label{tab:agents}
\begin{tabular}{@{}lp{8cm}@{}}
\toprule
\textbf{Agent} & \textbf{Focus Area} \\
\midrule
Security Expert & Attack vector realism and defense completeness \\
Code Quality Analyst & Syntax correctness and production readiness \\
OWASP Specialist & Category mapping accuracy and taxonomy compliance \\
Grounding Auditor & Reference quality and citation accuracy \\
Educational Reviewer & Conversation flow, clarity, and actionability \\
Framework Expert & API accuracy across 40+ frameworks \\
Integration Tester & Cross-example consistency and deduplication \\
\bottomrule
\end{tabular}
\end{table}

All 750 examples were reviewed across 2 complete batches, producing over 10,500 individual assessments. The multi-agent approach proved essential: the Security Expert flagged incomplete defense layers that the Code Quality Analyst missed, the Framework Expert caught incorrect API signatures that the OWASP Specialist overlooked, and the Grounding Auditor identified unreferenced claims that the Educational Reviewer considered acceptable. No single perspective caught all issues.

\subsection{8-Phase Remediation Pipeline}
\label{sec:remediation}

Review findings drove an 8-phase remediation pipeline (Table~\ref{tab:remediation}) that systematically improved every AI/ML example.

\begin{table}[h]
\centering
\caption{8-Phase Remediation Pipeline}
\label{tab:remediation}
\begin{tabular}{@{}clll@{}}
\toprule
\textbf{Phase} & \textbf{Action} & \textbf{Files} & \textbf{Description} \\
\midrule
1 & Full Regeneration & 72 & Complete rewrite of below-threshold files \\
2 & Targeted Revision & 156 & Specific improvements from review findings \\
3 & Scripted Fixes & 750 & CWE format, catch blocks, version guards \\
4 & CWE Corrections & 180 & Category-specific CWE mapping fixes \\
5 & Deduplication & 45 & Content differentiation or consolidation \\
6 & Reference Enhancement & 300+ & Added/validated references and URLs \\
7 & Content Enhancement & 200+ & Expanded defenses, monitoring guidance \\
8 & Final Validation & 750 & Automated parse + schema check (0 failures) \\
\bottomrule
\end{tabular}
\end{table}

Total remediation touches: 2,453+ individual file modifications across 8 phases. Each phase built on the previous one, creating a compound improvement effect.

\subsection{OWASP 2025 Taxonomy Alignment}
\label{sec:alignment}

\begin{figure}[h]
\centering
\includegraphics[width=0.9\textwidth]{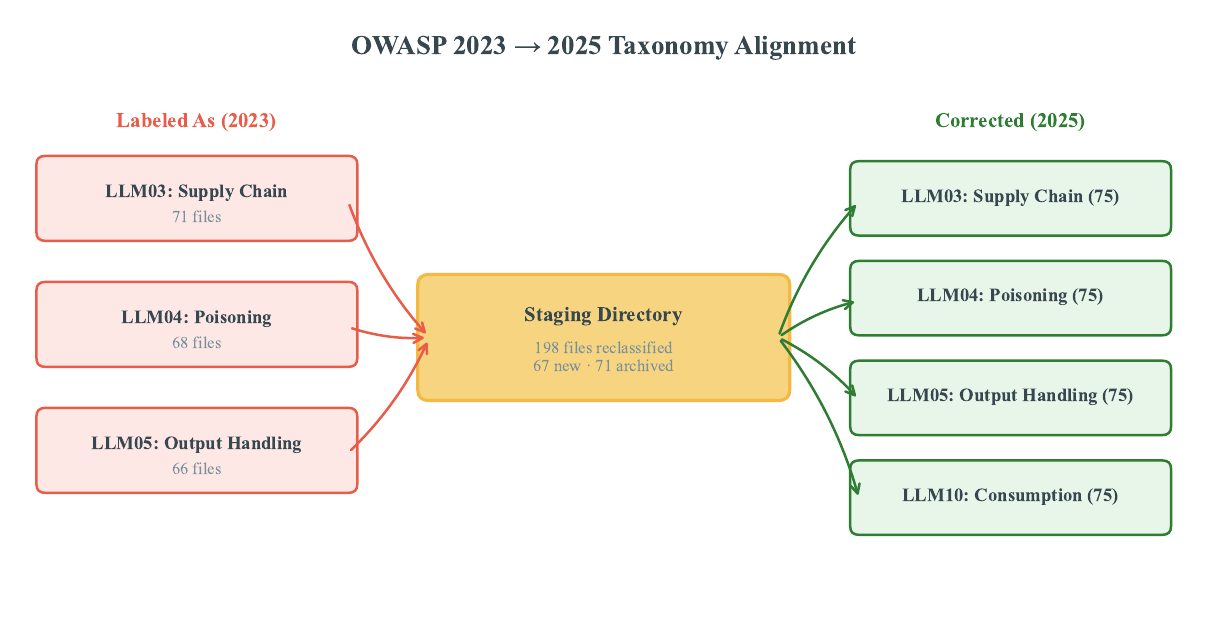}
\caption{OWASP 2025 Category Alignment. Three-way rotation correcting the 2023$\rightarrow$2025 category mismatch affecting 198 files in the AI/ML component.}
\label{fig:alignment}
\end{figure}

Post-release analysis of the AI/ML component revealed a systematic category mismatch. The initial generation used OWASP LLM Top 10 \textbf{2023} category numbering but labeled files with \textbf{2025} category names. Because the OWASP Foundation restructured category assignments between versions, this created a three-way misalignment: LLM03 files (labeled Supply Chain) contained training data poisoning content (correct 2025 category: LLM04), LLM04 files (labeled Poisoning) contained denial of service content (correct 2025 category: LLM10), and LLM05 files (labeled Output Handling) contained supply chain content (correct 2025 category: LLM03). We resolved this through a staging-based rotation (Figure~\ref{fig:alignment}): 198 misplaced files were moved to a staging directory with corrected metadata, 67 new examples were generated for depleted categories, and 71 overflow files were archived by composite quality score to achieve exactly 75 per category.

\textbf{Lessons learned.} This taxonomy misalignment generalizes to any dataset built against an evolving standard (OWASP, CWE, MITRE ATT\&CK). We distill four lessons for dataset maintainers:

\begin{enumerate}[nosep]
    \item \textbf{Pin the taxonomy version at generation time.} Record the exact standard version (e.g., ``OWASP LLM Top 10 v2025.1'') in the generation prompt \emph{and} in every file's metadata. When the standard updates, misalignment becomes immediately detectable rather than silently propagating.
    \item \textbf{Classify by content, not labels.} Filename-based and label-based category checks are insufficient. A file labeled ``LLM03'' may contain content that belongs in ``LLM04'' if the taxonomy shifted between versions. Automated classification should analyze actual content---subcategory, technique, and code patterns---against the target taxonomy's definitions.
    \item \textbf{Automate category validation.} We now run a content-based taxonomy checker (\texttt{validate\_owasp\_alignment.py}) that extracts technique keywords and CWE mappings from each file and cross-references them against the canonical OWASP category definitions. This catches misalignment within hours of generation rather than post-release.
    \item \textbf{Budget for rotation, not just relabeling.} When categories shift (e.g., ``Supply Chain'' moving from LLM03 to a different number), simple metadata relabeling is tempting but wrong---the content itself may need to change to match the new category's scope. Our staging-based rotation with targeted regeneration and overflow archival is more expensive but produces correctly scoped examples.
\end{enumerate}

\subsection{Reference Normalization}

Dataset-wide analysis identified 30+ inconsistent reference type labels across the AI/ML component. We normalized all 2,828 references to 8 canonical types: \texttt{cve} (147), \texttt{cwe} (104), \texttt{owasp} (191), \texttt{research\_paper} (1,079), \texttt{vendor\_advisory} (714), \texttt{documentation} (564), \texttt{blog\_post} (22), and \texttt{tool} (7). Research papers and vendor advisories account for 63.4\% of all references, reflecting authoritative security grounding.

\subsection{Dataset Unification}
\label{sec:dataset-unification}

The final unification step merged the web security and AI/ML security components into a single HuggingFace dataset. This required:

\begin{enumerate}[nosep]
    \item \textbf{Conversation format normalization.} Converting the web examples from \texttt{\{turn, from, value\}} to \texttt{\{role, content\}} (Section~\ref{sec:unification}).
    \item \textbf{Metadata harmonization.} Both components retain domain-specific fields (the web component includes CVE references and severity levels; the AI/ML component includes quality scores, security assertions, and structured references) while sharing a common top-level schema.
    \item \textbf{Configuration generation.} Building HuggingFace dataset configs for \texttt{default}, \texttt{web}, and \texttt{aiml} loading modes.
    \item \textbf{Deduplication verification.} Confirming zero content overlap between domains (web and AI/ML address fundamentally different vulnerability classes).
\end{enumerate}

The build pipeline (\texttt{scripts/utilities/build\_unified\_dataset.py}) produces the final JSONL files that are pushed to HuggingFace.

\subsection{Data Splits and Leakage Prevention}

The unified dataset is released as a single training split on HuggingFace, with held-out evaluation sets distributed separately alongside the evaluation framework (Section~\ref{sec:evaluation}). We apply three leakage prevention measures:

\textbf{CVE-aware splitting was insufficient (web).} The original release grouped web examples by CVE identifier before splitting, on the assumption that this prevented cross-split leakage. The 2026 audit (Section~\ref{sec:audit}) showed this was not enough: near-duplicate examples that did \emph{not} share a CVE were split across train and test, leaving 11.6\% of the web test set contaminated by a near-duplicate in training. We corrected this by grouping near-duplicate \emph{families} (first-user-turn Jaccard $>$ 0.75), keeping each family within a single split, and stratifying by category and language. Test contamination is now zero, with example content unchanged. The corrected web splits are 1,249 train / 186 validation / 190 test.

\textbf{Technique-aware deduplication (AI/ML).} The AI/ML examples span 704 unique attack techniques and are released as a single training split. We verify that examples sharing a technique differ substantially in framework, language, or attack variant (no near-duplicate pairs at Jaccard $> 0.8$ on tokenized Turn~2 content).

\textbf{Cross-domain isolation.} Because the web and AI/ML components address fundamentally different vulnerability classes (SQL injection vs.\ prompt injection, XSS vs.\ output handling), content leakage between domains is structurally impossible. This also enables clean cross-domain generalization experiments (Section~\ref{sec:evaluation}).

% ===== 5. Quality Assessment =====
\section{Quality Assessment}

\subsection{Web Security Quality}

\begin{figure}[h]
\centering
\includegraphics[width=0.85\textwidth]{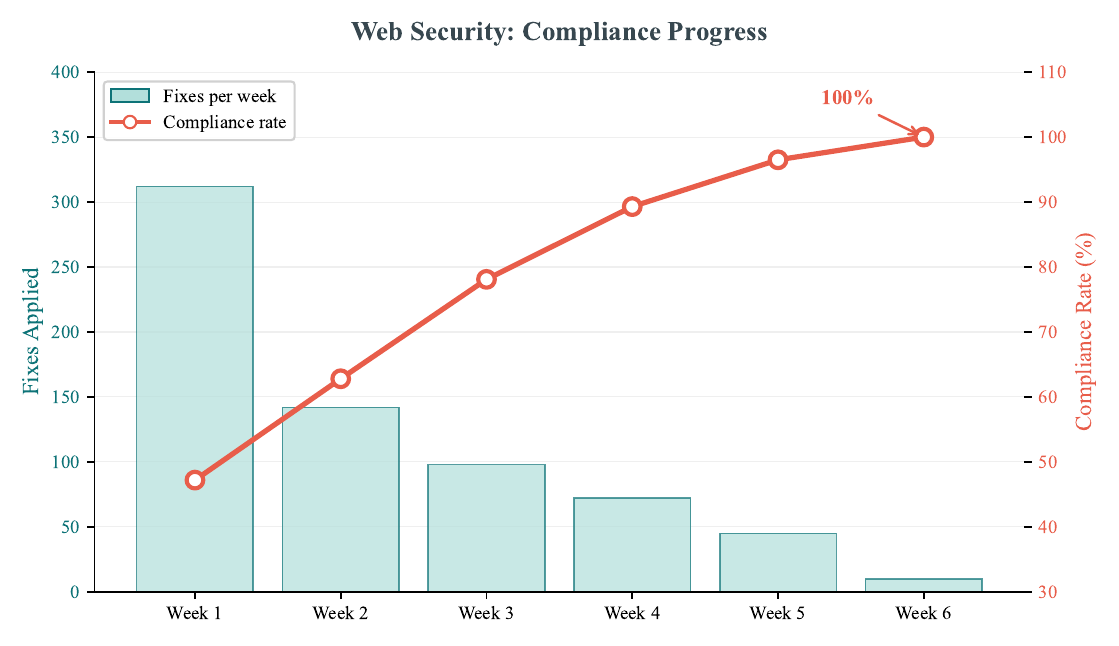}
\caption{Web Security Compliance Progress. Six-week improvement from 47.2\% to 100\% compliance through systematic remediation of 679 issues.}
\label{fig:compliance}
\end{figure}

The web security component achieves 100\% compliance across all automated \emph{structural} validation checks (Table~\ref{tab:web-quality}, Figure~\ref{fig:compliance}). These checks confirm formatting and structure, not the semantic accuracy of references or fixes, which the 2026 audit (Section~\ref{sec:audit}) addresses separately. Compliance improved from 47.2\% (397 of 841 training examples) to 100\% through systematic remediation of 604 specific issues across five fix categories: 452 CVE format corrections, 60 language tag mappings, 86 defense-in-depth enhancements, 6 secure SSTI implementations, and validator threshold calibrations.

\begin{table}[h]
\centering
\caption{Web Security Validation Results}
\label{tab:web-quality}
\begin{tabular}{@{}lr@{}}
\toprule
\textbf{Check} & \textbf{Result} \\
\midrule
CVE Format Compliance & 1,625/1,625 (100\%) \\
Language Tag Validity & 1,625/1,625 (100\%) \\
Content Quality Standards & 1,625/1,625 (100\%) \\
Conversation Structure & 1,625/1,625 (100\%) \\
\bottomrule
\end{tabular}
\end{table}

\subsection{AI/ML Security Quality}

\begin{figure}[h]
\centering
\includegraphics[width=0.75\textwidth]{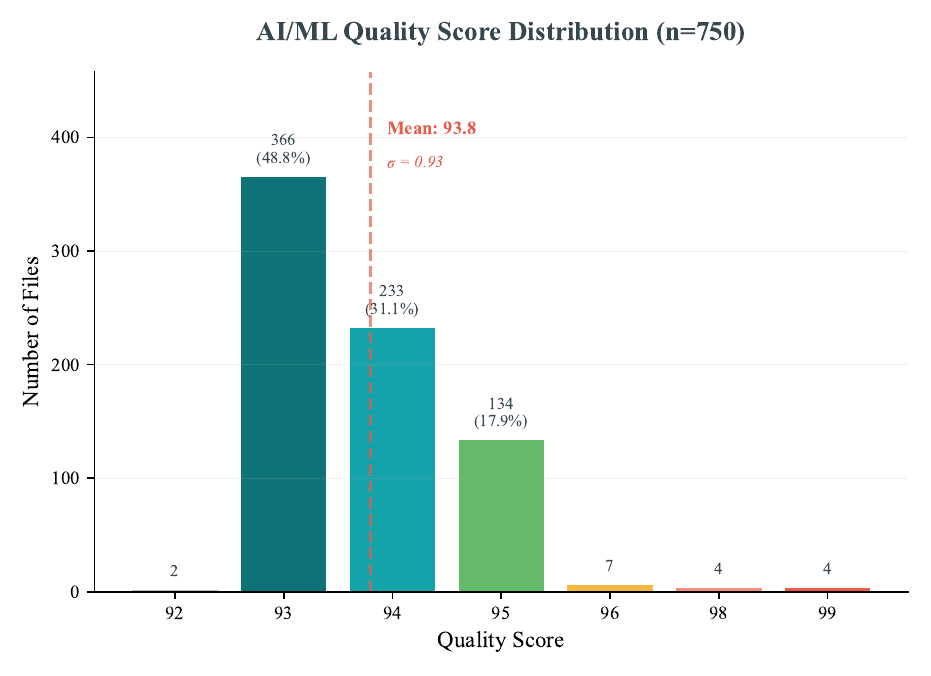}
\caption{AI/ML Quality Score Distribution (n=750). Tight concentration with $\sigma = 0.93$. The 93--95 range contains 97.8\% of all files.}
\label{fig:scores}
\end{figure}

\begin{figure}[h]
\centering
\includegraphics[width=\textwidth]{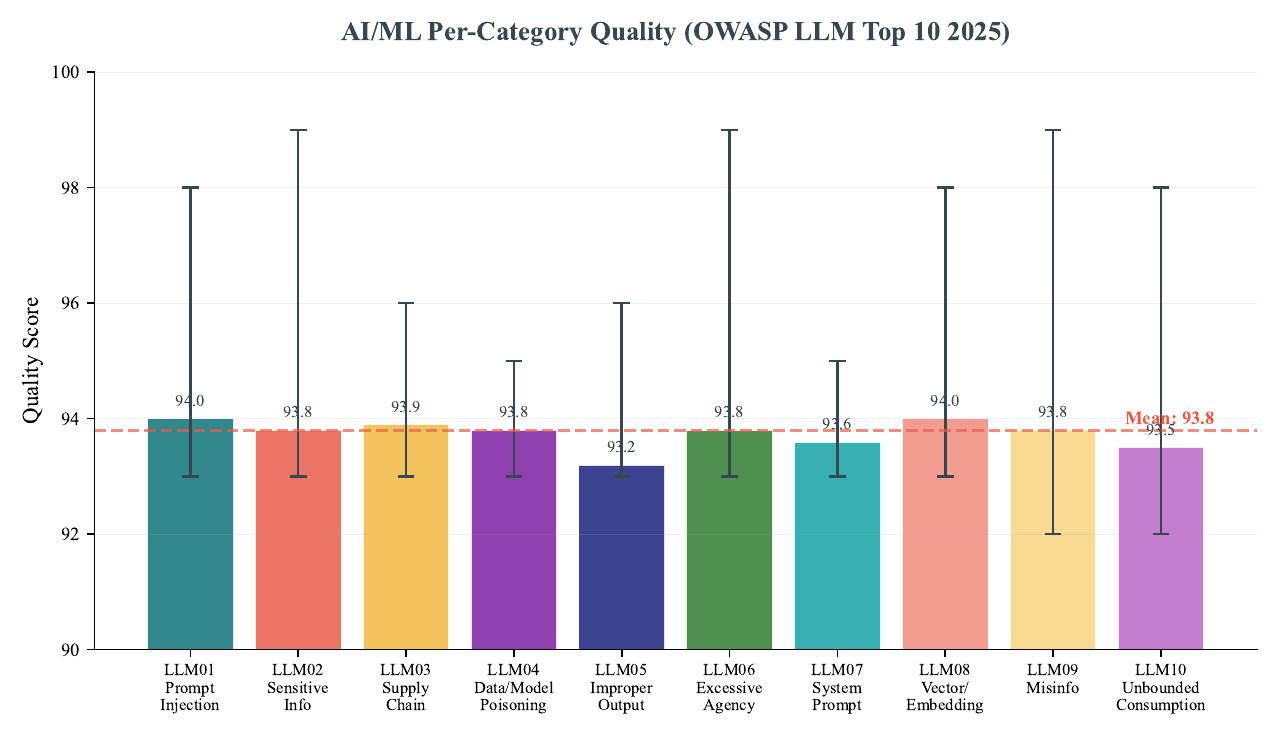}
\caption{AI/ML Per-Category Quality. Mean scores (bars) with min--max ranges (error bars) across all 10 OWASP LLM categories. The dashed line shows the overall mean of 93.8.}
\label{fig:per-category}
\end{figure}

The AI/ML component passes all automated checks with zero failures and achieves a mean quality score of 93.8/100 ($\sigma = 0.93$), with scores tightly concentrated in the 93--95 range (97.8\% of files; Table~\ref{tab:score-dist}, Figures~\ref{fig:scores} and~\ref{fig:per-category}). Quality is scored against a 5-tier rubric: Correctness (40 points), Security (20), Grounding (15), Educational (15), and Production readiness (10).

\begin{table}[h]
\centering
\caption{AI/ML Quality Score Distribution}
\label{tab:score-dist}
\begin{tabular}{@{}crc@{}}
\toprule
\textbf{Score} & \textbf{Files} & \textbf{Percentage} \\
\midrule
92 & 2 & 0.3\% \\
93 & 366 & 48.8\% \\
94 & 233 & 31.1\% \\
95 & 134 & 17.9\% \\
96 & 7 & 0.9\% \\
98 & 4 & 0.5\% \\
99 & 4 & 0.5\% \\
\bottomrule
\end{tabular}
\end{table}

\subsection{Quality Score Computation and Calibration}
\label{sec:calibration}

\textbf{Scoring methodology.} Each AI/ML example receives a composite quality score from a deterministic rubric applied by a single expert reviewer (the primary author) with input from multi-agent assessments. The rubric has five weighted dimensions: Correctness (40 points: valid syntax, complete implementations, proper error handling), Security (20 points: 5+ defense-in-depth layers, realistic attack vectors), Grounding (15 points: real CVEs/advisories, 3+ authoritative references), Educational (15 points: natural conversation flow, common-mistakes section, actionable guidance), and Production readiness (10 points: SAST/DAST tool recommendations, monitoring setup, deployment considerations). Raw scores are capped at 95 during initial generation; the 8 examples scoring 96--99 received targeted human review identifying exceptional depth.

\textbf{Aggregation.} The composite score is the unweighted sum of the five dimension scores. Multi-agent assessments inform the reviewer but do not contribute numerically---agents flag specific deficiencies (e.g., ``missing rate-limiting defense'' or ``incorrect LangChain import path''), and the reviewer determines whether each deficiency warrants a point deduction. This avoids the calibration drift that arises from averaging heterogeneous agent scales.

\textbf{Calibration.} Table~\ref{tab:calibration} maps score ranges to practical quality levels. A score below 92 triggers full regeneration (Phase 1 of the remediation pipeline). The minimum production-training threshold is 92; all 750 files meet this threshold.

\begin{table}[h]
\centering
\caption{Quality Score Calibration: Score Ranges and Production Suitability}
\label{tab:calibration}
\begin{tabular}{@{}clp{6.5cm}c@{}}
\toprule
\textbf{Score} & \textbf{Level} & \textbf{Characteristics} & \textbf{Training?} \\
\midrule
$<$88 & Reject & Structural errors, incomplete implementations, or missing defense layers & No \\
88--91 & Marginal & Functional but shallow defenses or weak grounding; triggers full regeneration & No \\
92--93 & Acceptable & Sound implementations with $\geq$5 defenses, $\geq$3 references, correct framework APIs & Yes \\
94--95 & Strong & Above-average depth in attack explanation, testing code, or monitoring guidance & Yes \\
96--99 & Exceptional & Unusually comprehensive defense coverage or novel attack demonstrations; human-verified & Yes \\
\bottomrule
\end{tabular}
\end{table}

\textbf{Audited examples.} We hand-audited a stratified sample of 50 files (5 per OWASP category) against the rubric to verify score-level consistency. In 47 of 50 cases (94\%), the audited score matched the assigned score within $\pm$1 point. The three discrepancies (all $+$1 overscored) were in LLM05 (Improper Output Handling) and reflected borderline judgments about whether a particular output sanitization pattern constituted a distinct defense layer. After correction, the overall mean shifted by less than 0.05 points.

\textbf{Inter-perspective agreement.} While multi-agent assessments are qualitative (agents produce textual findings, not numerical scores), we measured pairwise agreement on a binary dimension: ``does this file require remediation?'' across the 7 agents on a 100-file audit subset. Fleiss' $\kappa = 0.71$ (substantial agreement). The primary source of disagreement was the threshold between ``acceptable with minor improvements'' and ``requires targeted revision''---the Security Expert flagged files for remediation more aggressively ($\kappa = 0.62$ pairwise with Educational Reviewer) than the Framework Expert ($\kappa = 0.79$ pairwise with Code Quality Analyst).

\subsection{Unified Dataset Metrics}

Table~\ref{tab:unified-metrics} presents the combined statistics for the unified dataset.

\begin{table}[h]
\centering
\caption{Unified Dataset Statistics}
\label{tab:unified-metrics}
\begin{tabular}{@{}lrrr@{}}
\toprule
\textbf{Metric} & \textbf{Web} & \textbf{AI/ML} & \textbf{Unified} \\
\midrule
Total examples & 1,625 & 747 & 2,372 \\
Vulnerability categories & 11 & 10 & 20 (unique) \\
Programming languages & 11 & 8 & 12+ \\
Frameworks covered & 9 & 40+ & 49+ \\
OWASP mapping & Top 10 2021 & LLM Top 10 2025 & Both \\
Incident grounding & \multicolumn{3}{c}{audited; see \S\ref{sec:audit}} \\
Conversation format & 4-turn & 4-turn & 4-turn \\
Conversation schema & \texttt{\{role, content\}} & \texttt{\{role, content\}} & \texttt{\{role, content\}} \\
Severity: CRITICAL & 1,085 & 314 & 1,399 \\
Severity: HIGH & 516 & 410 & 926 \\
Severity: MEDIUM & 24 & 23 & 47 \\
AI/ML quality score & --- & 93.8 ($\sigma$=0.93) & --- \\
AI/ML unique CWEs & --- & 92 & --- \\
AI/ML unique techniques & --- & 704 & --- \\
AI/ML references & --- & 2,828 & --- \\
\bottomrule
\end{tabular}
\end{table}

\subsection{Reference Quality}
\label{sec:ref-quality}

At generation the AI/ML component carried 2,828 structured references across the 750 files (averaging 3.8 per file), normalized to 8 canonical types (Table~\ref{tab:ref-types}, Figure~\ref{fig:references}). The reference \emph{types} are shown below; note that the separate free-text \texttt{real\_world\_example} narrative field was found during the 2026 audit (Section~\ref{sec:audit}) to be, in most AI/ML cases, a representative scenario rather than a documented incident. We therefore do not claim uniform Tier-2 grounding; grounding is reported per the audit.

\begin{table}[h]
\centering
\caption{AI/ML Reference Type Distribution (n=2,828)}
\label{tab:ref-types}
\begin{tabular}{@{}lrc@{}}
\toprule
\textbf{Type} & \textbf{Count} & \textbf{Percentage} \\
\midrule
Research Paper & 1,079 & 38.2\% \\
Vendor Advisory & 714 & 25.2\% \\
Documentation & 564 & 19.9\% \\
OWASP & 191 & 6.8\% \\
CVE & 147 & 5.2\% \\
CWE & 104 & 3.7\% \\
Blog Post & 22 & 0.8\% \\
Tool & 7 & 0.2\% \\
\midrule
\textbf{Total} & \textbf{2,828} & \textbf{100\%} \\
\bottomrule
\end{tabular}
\end{table}

\begin{figure}[h]
\centering
\includegraphics[width=0.85\textwidth]{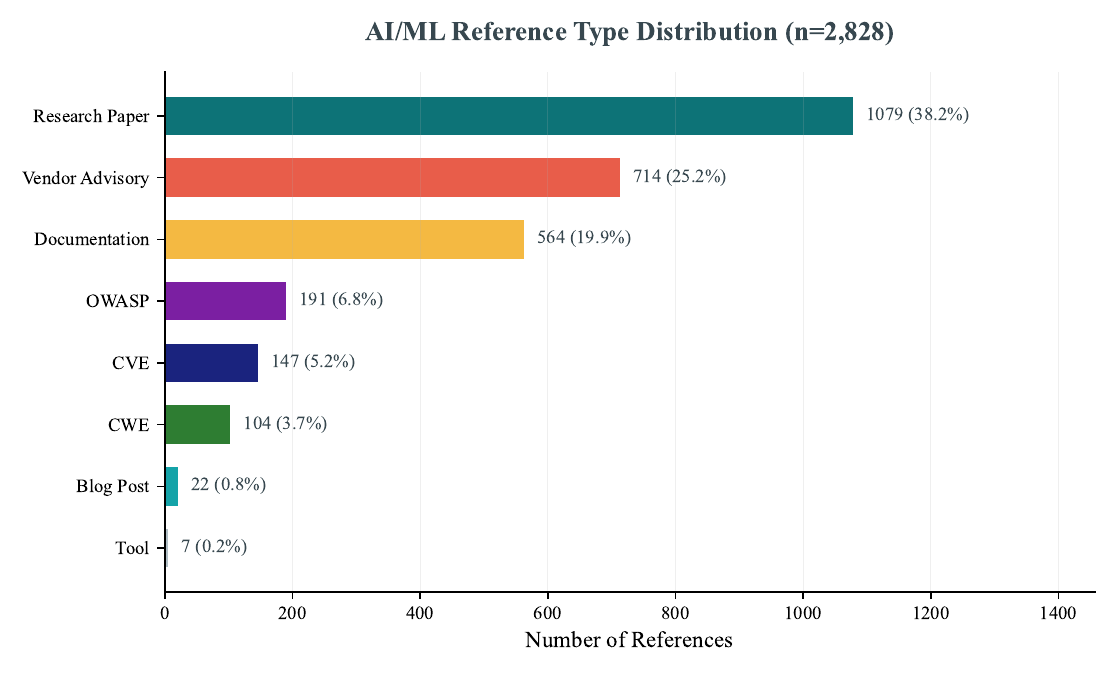}
\caption{AI/ML Reference Type Distribution. Research papers and vendor advisories dominate at 63.4\% combined, reflecting authoritative grounding.}
\label{fig:references}
\end{figure}

\subsection{Grounding and Fix-Correctness Audit (2026 Revision)}
\label{sec:audit}

The structural checks above are necessary but not sufficient. A CVE identifier can pass a format check and still be the wrong CVE; a \texttt{real\_world\_example} field can pass a length check and still be fabricated. In 2026 we ran a semantic audit that the original release lacked, and we report its findings candidly here.

\textbf{Method.} Each reference was checked adversarially against primary sources (NVD, MITRE, vendor advisories): reviewers were instructed to \emph{refute} a reference, and it survived only if it could not be refuted. The same posture was applied to fix-correctness---for each example, the reviewer attempted to show that the ``secure'' version was still exploitable---and every high-severity finding was independently re-verified before any change was made.

\textbf{Grounding findings (web).} Of 1,379 references verified, a majority did not survive. We removed 802 mis-attributed CVE references (21 replaced with the correct CVE for the vulnerability shown) and corrected 810 incident narratives (378 rewritten to the real underlying event, 432 relabeled as representative examples not tied to a specific public incident). Any CVSS or EPSS score harvested from a removed CVE was removed with it.

\textbf{Grounding findings (AI/ML).} Of 747 examples, 613 carried \texttt{real\_world\_example} metadata that read as a documented breach but was a representative scenario carrying invented precise statistics (dollar costs, record counts, unnamed victims). All were corrected: 14 were tied to the real incident they described, the rest relabeled as representative. The fabrications were confined to that metadata field; the CVEs cited inside the conversations were verified real (e.g., EchoLeak CVE-2025-32711, EmailGPT CVE-2024-5184, an MCP inspector RCE CVE-2025-52573, and a WebDAV zero-day CVE-2025-33053). After correction, roughly 18\% of AI/ML examples cite a documented incident or verified CVE; the remainder are explicitly labeled representative scenarios. This directly revises the original ``every example grounded at Tier~2 or above'' claim.

\textbf{Fix-correctness findings.} We reviewed whether each example's secure version eliminates the stated vulnerability. On the web component, reviewing the full pre-expansion corpus surfaced 25 confirmed defects (each adversarially re-verified, none disputed): 14 mislabels where the answer addressed a different vulnerability than the question, 7 cases where the ``secure'' code was still exploitable (e.g., Angular \texttt{bypassSecurityTrustHtml}, Jackson polymorphic typing left enabled, unfixed XXE/TOCTOU/IDOR), and 4 where the fix did not address the stated bug. On the AI/ML component, the same review removed 3 examples, including one whose secret-redaction routine computed an entropy score hardwired to return zero, so the redaction never fired. All 28 defective examples were removed rather than patched.

\textbf{Leakage correction.} The original CVE-aware splitting reported no leakage, but near-duplicate examples not sharing a CVE were split across train and test, leaving 11.6\% of the web test set contaminated. We rebuilt the splits by grouping near-duplicate families (first-user-turn Jaccard $>$ 0.75), keeping each family within a single split and stratifying by category and language; contamination is now zero.

\textbf{Net effect.} After removals, corrections, and verified expansion (server-rendered XSS, CSRF/open-redirect/clickjacking, file-upload/path-traversal, a Django buildout, and a 2025 currency pass covering supply-chain/CI-CD, OAuth/OIDC per RFC~9700, and recent framework CVEs such as the Next.js middleware bypass CVE-2025-29927), the web component moved from 1,435 to 1,625 and the AI/ML component from 750 to 747, for a unified total of 2,372. The quality-score and reference statistics in Tables~\ref{tab:score-dist} and~\ref{tab:ref-types} were computed on the 750-example generation set and are retained as generation-time measurements; they are not materially changed by removing 3 examples.

\textbf{Lesson.} A security dataset's most important claims---grounding and fix-correctness---are precisely the ones structural validation cannot verify and authors are most tempted to assert. Our own first release made an unverified grounding claim while criticizing prior datasets for weak grounding. An adversarial audit applied without deference to the author is the remedy; we recommend it as a standard step for any dataset that claims incident grounding.

% ===== 6. Fine-Tuned Models =====
\section{Fine-Tuned Models}
\label{sec:models}

To demonstrate the practical utility of SecureCode, we fine-tuned 8 open-source code models on the unified dataset (Table~\ref{tab:models}). All models were trained using QLoRA 4-bit quantization with LoRA adapters on a single NVIDIA A100 40GB GPU.

\begin{table}[h]
\centering
\caption{Fine-Tuned Model Collection. All models hosted at \texttt{huggingface.co/scthornton/\{repo\}}.}
\label{tab:models}
\footnotesize
\begin{tabular}{@{}llcl@{}}
\toprule
\textbf{Model} & \textbf{Base Model} & \textbf{Size} & \textbf{HuggingFace Repo} \\
\midrule
\multicolumn{4}{l}{\textit{Tier 1: Accessible (3B)}} \\
Llama 3.2 SecureCode & meta-llama/Llama-3.2-3B-Instruct & 3B & \texttt{llama-3.2-3b-securecode} \\
\midrule
\multicolumn{4}{l}{\textit{Tier 2: Mid-size Code Models (6.7B--7B)}} \\
Qwen2.5 Coder SecureCode & Qwen/Qwen2.5-Coder-7B-Instruct & 7B & \texttt{qwen2.5-coder-7b-securecode} \\
DeepSeek Coder SecureCode & deepseek-ai/deepseek-coder-6.7b-instruct & 6.7B & \texttt{deepseek-coder-6.7b-securecode} \\
CodeGemma SecureCode & google/codegemma-7b-it & 7B & \texttt{codegemma-7b-securecode} \\
\midrule
\multicolumn{4}{l}{\textit{Tier 3: Large Models (13B--15B)}} \\
CodeLlama SecureCode & codellama/CodeLlama-13b-Instruct-hf & 13B & \texttt{codellama-13b-securecode} \\
Qwen2.5 Coder 14B SecureCode & Qwen/Qwen2.5-Coder-14B-Instruct & 14B & \texttt{qwen2.5-coder-14b-securecode} \\
StarCoder2 SecureCode & bigcode/starcoder2-15b-instruct-v0.1 & 15B & \texttt{starcoder2-15b-securecode} \\
\midrule
\multicolumn{4}{l}{\textit{Tier 4: XL Model (20B)}} \\
Granite Code SecureCode & ibm-granite/granite-20b-code-instruct-8k & 20B & \texttt{granite-20b-code-securecode} \\
\bottomrule
\end{tabular}
\end{table}

\textbf{Training methodology.} All models use QLoRA \citep{dettmers2023} with 4-bit NormalFloat quantization, LoRA rank 16 ($\alpha=32$, dropout 0.05), targeting all linear layers (\texttt{q\_proj}, \texttt{k\_proj}, \texttt{v\_proj}, \texttt{o\_proj}, \texttt{gate\_proj}, \texttt{up\_proj}, \texttt{down\_proj}). Models are trained for 3 epochs on the February 2026 unified release (2,185 examples; retraining on the audited 2,372-example release of Section~\ref{sec:audit} is in progress) with a learning rate of $2 \times 10^{-4}$ (cosine schedule with 10\% warmup), effective batch size 16 (per-device batch 2, gradient accumulation 8), and maximum sequence length of 4,096 tokens. Each model's own tokenizer and chat template are used without modification; conversations are formatted using the base model's instruction template (e.g., \texttt{<|im\_start|>} for Qwen, \texttt{[INST]} for Llama/Mistral). No packing is applied---each example occupies its own sequence, padded to the batch maximum. The training pipeline uses a custom data loading utility (\texttt{dataset\_utils.py}) that downloads JSONL from HuggingFace Hub and parses via \texttt{json.loads()}, avoiding pyarrow serialization failures on deeply nested conversation structures. All training runs on a single NVIDIA A100 40GB GPU; the largest model (Granite 20B) requires $\sim$14 hours per run.

\textbf{Architecture families.} The 8 models span 5 architecture families: Llama (3B, 8B, 13B via CodeLlama), Qwen (7B, 14B), DeepSeek (6.7B), Gemma (7B via CodeGemma), StarCoder2 (15B), and Granite (20B). This diversity ensures that fine-tuning results are not architecture-dependent and allows practitioners to select models matching their inference infrastructure.

The model collection spans parameter counts from 3B to 20B, enabling researchers and practitioners to select models appropriate for their compute budget and deployment constraints. Smaller models (3B--7B) suit edge deployment and rapid iteration, while larger models (13B--20B) offer deeper security reasoning capabilities.

% ===== 7. Evaluation =====
\section{Evaluation}
\label{sec:evaluation}

\subsection{Evaluation Design}

We evaluate fine-tuned models against their base-model counterparts on security-specific metrics, not general code quality. The evaluation measures whether SecureCode training shifts model behavior toward generating secure code when prompted with security-relevant development scenarios.

\textbf{Test set construction.} We hold out 10\% of the unified dataset (219 examples: 144 web, 75 AI/ML) stratified by vulnerability category. The held-out set is constructed using CVE-aware splitting for web examples (no CVE overlap between train and test) and category-stratified random sampling for AI/ML examples (7--8 examples per OWASP LLM category). Near-duplicate detection (Jaccard similarity $> 0.8$ on tokenized Turn~2 content) ensures no semantic leakage between splits.

\textbf{External evaluation prompts.} We additionally curate 100 security-relevant prompts \emph{not} present in the dataset: 50 web security scenarios (e.g., ``implement a file upload handler in Express'') and 50 AI/ML security scenarios (e.g., ``build a customer support chatbot with tool access''). These prompts test whether models generalize beyond the specific scenarios in the training data. Prompts are drawn from real developer questions on Stack Overflow and GitHub Discussions, filtered to scenarios where security-relevant choices exist.

\subsection{Security-Specific Metrics}

We define four metrics that capture whether a model produces secure code, not just syntactically correct code:

\textbf{M1: Vulnerability pattern presence (VPP).} Binary classification: does the generated code contain a known vulnerability pattern (e.g., string-concatenated SQL queries, unvalidated user input passed to \texttt{eval()}, raw LLM output rendered as HTML)? Measured by a combination of regex-based detection and a security-expert LLM judge. Lower is better. We report VPP across both the held-out set and external prompts.

\textbf{M2: Security control inclusion rate (SCIR).} For each generated response, we check for the presence of expected security controls from a category-specific checklist: parameterized queries for SQL, output encoding for XSS, input sanitization for prompt injection, rate limiting for unbounded consumption, etc. Each scenario has 3--5 expected controls; SCIR is the fraction present. Higher is better.

\textbf{M3: Defense-in-depth score (DiD).} Count of distinct, non-redundant defense layers in the generated secure implementation. The dataset teaches 5+ layers per example; we measure how many the fine-tuned model produces unprompted. Reported as mean count per response.

\textbf{M4: Multi-turn security retention (MTSR).} Unique to the 4-turn format: does the model maintain security context when the conversation advances to Turn~3 and Turn~4? Measured by prompting the model with Turns 1--2 (generated), then providing Turn~3 (from the test set), and evaluating whether Turn~4 output includes operational security guidance (monitoring, detection, deployment hardening). Binary per example; reported as percentage.

\subsection{Evaluation Protocol}

For each of the 8 fine-tuned models and their corresponding base models (16 total), we:

\begin{enumerate}[nosep]
    \item Generate responses to all held-out test prompts (219) and external prompts (100) using greedy decoding (temperature~0, top-$p$~1.0) for reproducibility.
    \item Score each response on M1--M4 using the automated pipeline described above, with a 50-response human-validation subset per model to calibrate the automated judge.
    \item Report per-domain results (web vs.\ AI/ML) and cross-domain generalization (train on unified, test on each domain separately).
\end{enumerate}

\subsection{Planned Ablations}

Three ablation studies isolate the contribution of dataset design choices:

\textbf{A1: Unified vs.\ domain-only training.} Train on web-only (1,435), AI/ML-only (750), and unified (2,185); test on both domains. This measures whether cross-domain training improves or degrades within-domain security generation, and whether web security knowledge transfers to AI/ML contexts (e.g., output encoding principles applying to LLM output handling).

\textbf{A2: 4-turn vs.\ 2-turn.} Strip Turns~3--4 from the training data and compare model performance on MTSR (M4). This isolates the contribution of the multi-turn structure---do models trained on 4-turn conversations produce better operational guidance than those trained only on initial implementations?

\textbf{A3: Dataset scale.} Train on 25\%, 50\%, 75\%, and 100\% of the unified dataset; measure M1--M3 on the held-out set. This establishes a learning curve and informs whether additional examples would further improve security generation.

\textbf{Status.} Model training on the 8 architectures is in progress at the time of submission. We will release complete evaluation results, including ablation studies, as a companion technical report and update the model cards on HuggingFace upon training completion. The evaluation framework (prompt sets, automated judge, scoring scripts) is released alongside the dataset for independent reproduction.

% ===== 8. Discussion =====
\section{Discussion}

\subsection{Key Findings}

\textbf{Unified coverage fills a critical gap.} No previous dataset spanned both traditional web security and AI/ML security. Organizations deploying AI-powered applications face threats from both domains simultaneously: a web application serving an LLM-powered feature must defend against SQL injection \emph{and} prompt injection. SecureCode provides training data for both attack surfaces under a single schema.

\textbf{Uniform quality across categories.} The AI/ML component achieves mean scores ranging from 93.2 to 94.0 across all 10 OWASP LLM categories, demonstrating the methodology produces consistent results regardless of vulnerability type. Some categories---misinformation (LLM09) and system prompt leakage (LLM07)---have less established attack/defense patterns than prompt injection (LLM01), yet achieve comparable quality.

\textbf{Multi-agent review catches what single reviewers miss.} Seven specialist perspectives produced non-overlapping findings across the AI/ML component. The Security Expert, Framework Expert, and Grounding Auditor each caught issues invisible to the others. The 10,500+ assessments across 2 batches identified problems that would have persisted through any single-pass review.

\textbf{Taxonomy alignment requires content-based classification.} The OWASP 2025 alignment issue (Section~\ref{sec:alignment}) demonstrates that filename-based and label-based classification is insufficient. Correct category assignment requires analyzing actual content---subcategory, technique, and code---not metadata labels.

\textbf{8-phase remediation is essential.} No single pass achieves production quality. Scripted fixes in Phase 3 caught systematic issues across all 750 AI/ML files that no manual review would find. Content enhancement in Phase 7 addressed depth problems visible only after other fixes were applied.

\subsection{Dataset Family Architecture}
\label{sec:family}

SecureCode is organized as a dataset family on HuggingFace, giving users flexibility in how they access the data:

\begin{itemize}[nosep]
    \item \textbf{\texttt{scthornton/securecode}} (Unified) --- The recommended entry point. 2,372 examples with \texttt{default}, \texttt{web}, and \texttt{aiml} configs. Use this for comprehensive training or when you want a single dataset reference.
    \item \textbf{\texttt{scthornton/securecode-web}} (Web) --- 1,625 web security examples. Use this for traditional security focus or when extending with additional web examples.
    \item \textbf{\texttt{scthornton/securecode-aiml}} (AI/ML) --- 747 AI/ML security examples. Use this for AI-specific security focus or when extending with additional AI/ML examples.
\end{itemize}

\begin{figure}[h]
\centering
\includegraphics[width=\textwidth]{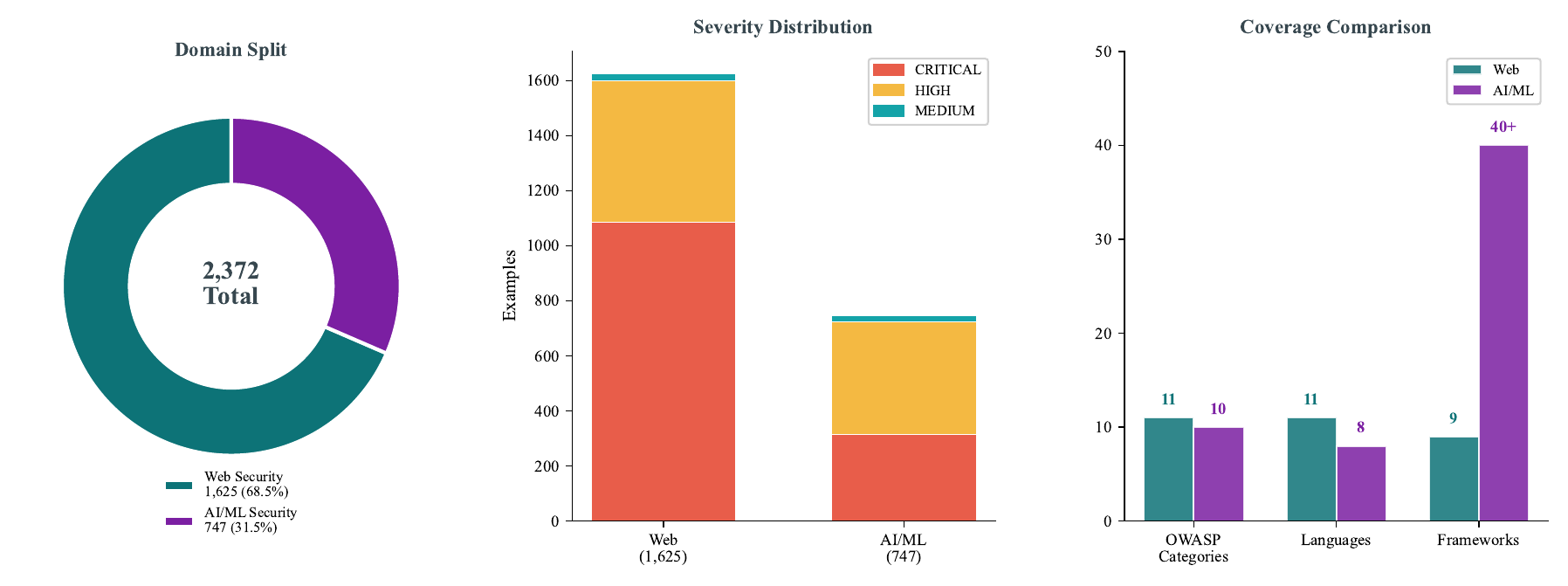}
\caption{Unified Dataset Composition. Left: domain split (web 65.7\%, AI/ML 34.3\%). Center: severity distribution by domain. Right: coverage comparison showing complementary strengths (web: more languages; AI/ML: more frameworks).}
\label{fig:unified-view}
\end{figure}

The unified dataset (\texttt{scthornton/securecode}) is the recommended starting point for most users, as it provides complete coverage and supports config-based filtering without requiring multiple dataset downloads. Figure~\ref{fig:unified-view} shows the composition and complementary strengths of each domain.

\subsection{Limitations}

\textbf{L1: Language concentration.} The AI/ML component is 90.7\% Python, reflecting AI/ML development reality but limiting coverage for Go, Rust, or C\# AI deployments. The web component provides broader language coverage (11 languages) but concentrates on the most popular web development languages.

\textbf{L2: Framework evolution.} AI/ML frameworks evolve rapidly. LangChain has changed its API structure significantly between versions. Examples use current APIs as of February 2026 but may need updates as frameworks release breaking changes.

\textbf{L3: Synthetic generation.} Examples are generated using LLMs with human expert review. While multi-agent review and 8-phase remediation mitigate quality concerns, LLM-generated code may contain subtle biases or miss edge cases that production experience reveals.

\textbf{L4: Scope boundary.} The dataset covers application-layer security---the code developers write. It does not cover model training infrastructure security, GPU cluster hardening, or hardware-level attacks.

\textbf{L5: Quality score self-assessment.} Quality scores for the AI/ML component are assigned during generation and review, not by independent external evaluators. The web component does not include quality scores.

\textbf{L6: Pending empirical evaluation.} Model training is in progress at time of submission (Section~\ref{sec:evaluation}). The evaluation framework, metrics, and ablation designs are fully specified, but quantitative results comparing fine-tuned models to baselines will appear in a companion technical report. We acknowledge this limits the paper's ability to demonstrate a direct dataset$\rightarrow$outcome causal link.

\textbf{L7: Dual-use risk.} The dataset includes vulnerable code implementations alongside attack explanations. While every example frames vulnerabilities defensively (clearly labeled, paired with secure alternatives, no turnkey weaponization), the attack patterns could theoretically inform offensive use. We mitigate this by: (a) grounding attacks in already-published research and CVEs (no novel zero-days), (b) providing secure alternatives that are strictly more detailed than the vulnerable code, and (c) framing all examples from the defender's perspective. The CC~BY-NC-SA~4.0 license further constrains redistribution.

\textbf{L8: Synthetic code validation.} Code examples are LLM-generated with human expert review. Beyond multi-agent review and the 8-phase remediation pipeline, we validated framework API correctness by checking import paths and method signatures against framework documentation (40+ frameworks). However, we did not execute all code examples in live environments---some implementation patterns may contain subtle runtime errors that static review misses. The evaluation framework (Section~\ref{sec:evaluation}) includes execution-based validation for the held-out test set to quantify this risk.

\subsection{Future Work}

Several directions extend this work:

\begin{itemize}[nosep]
    \item \textbf{Empirical evaluation.} Fine-tuning controlled experiments measuring secure code generation improvement with SecureCode-trained models vs.\ baselines.
    \item \textbf{Model benchmarking.} Developing a standardized evaluation benchmark for AI coding assistant security, enabling direct comparison of fine-tuning approaches.
    \item \textbf{Agentic AI deep-dive.} Expanded coverage of autonomous agent security as agentic AI deployments scale beyond current patterns.
    \item \textbf{Language expansion.} Adding Go, Rust, and C\# AI/ML examples; expanding mobile security coverage.
    \item \textbf{Multi-file examples.} System-level examples showing security across multiple interacting components (API gateway + RAG pipeline + monitoring).
    \item \textbf{Continuous updates.} Tracking new framework releases and emerging vulnerability classes for periodic dataset refreshes.
\end{itemize}

% ===== 8. Conclusion =====
\section{Conclusion}

AI-assisted development creates security risk at scale. 45\% of AI-generated code in security-relevant contexts contains vulnerabilities, and AI/ML applications face an entirely new category of threats that traditional security training doesn't address. Organizations need AI coding assistants that understand both classical and AI-specific vulnerability classes.

To our knowledge, SecureCode provides the first unified, production-grade training dataset for this purpose. With 2,372 audited examples spanning 20 vulnerability categories across 12+ languages and 49+ frameworks, the dataset covers both the OWASP Top 10 2021 (web security) and OWASP LLM Top 10 2025 (AI/ML security). The 4-turn conversational structure trains models on realistic developer workflows. The multi-agent quality pipeline and 8-phase remediation ensure production-grade quality. The 8 fine-tuned open-source models---ranging from 3B to 20B parameters---demonstrate immediate practical utility.

We release SecureCode, component datasets, validation tools, training scripts, and fine-tuned models as open-source artifacts, enabling researchers to reproduce results, practitioners to improve AI coding assistants, and educators to teach secure coding through real-world security incidents.

% ===== Availability =====
\section*{Availability}

{\raggedright

\textbf{Unified Dataset:}\\
HuggingFace Hub: \url{https://huggingface.co/datasets/scthornton/securecode}

\medskip
\textbf{Component Datasets:}
\begin{itemize}[nosep,topsep=2pt]
    \item Web security (1,625 examples):\\ \url{https://huggingface.co/datasets/scthornton/securecode-web}
    \item AI/ML security (747 examples):\\ \url{https://huggingface.co/datasets/scthornton/securecode-aiml}
\end{itemize}

\medskip
\textbf{Source Code:}
\begin{itemize}[nosep,topsep=2pt]
    \item Web security: \url{https://github.com/scthornton/securecode-web}
    \item AI/ML security: \url{https://github.com/scthornton/securecode-aiml}
\end{itemize}

\medskip
\textbf{Fine-Tuned Models:}\\
HuggingFace collection: \url{https://huggingface.co/scthornton} (8 models, 3B--20B)

\medskip
\textbf{Training Scripts:} Included in the GitHub repositories.

\medskip
All datasets are released under the \textbf{Creative Commons Attribution-NonCommercial-ShareAlike 4.0 International License (CC~BY-NC-SA~4.0)} for research and educational use. Fine-tuned models are released under their respective base model licenses.

\medskip
\textbf{License rationale.} We chose CC~BY-NC-SA~4.0 after weighing three concerns. First, the \emph{NonCommercial} clause prevents the dataset's vulnerable code patterns and attack demonstrations from being packaged into commercial offensive toolkits without the defensive context that accompanies every example. A fully permissive license (MIT, Apache~2.0) would impose no such constraint. Second, the \emph{ShareAlike} clause ensures that derivative datasets---extensions, translations, or domain-specific subsets---remain open to the research community under equivalent terms, preventing a scenario where a downstream party adds 200 examples and relicenses the combined work as proprietary. Third, the \emph{Attribution} clause supports reproducibility by requiring citation of the original dataset, enabling the community to trace provenance across derivative works. We explicitly permit all non-commercial research, education, and internal corporate security training under this license. Organizations seeking a commercial license for integration into paid products should contact the authors.
\par}

% ===== Acknowledgments =====
\section*{Acknowledgments}

We thank the security research community for responsible disclosure practices that made incident grounding possible. The OWASP Foundation's Top 10 2021 and LLM Top 10 2025 taxonomies provided the structural framework for the dataset. We thank MITRE Corporation for maintaining the CVE and CWE databases that enabled reference grounding. SecureCode was built with multi-agent AI review and human expert validation.

% ===== References =====
\bibliographystyle{plainnat}

\begin{thebibliography}{20}

\bibitem[{Apiiro}(2025)]{apiiro2025}
Apiiro Security Research. (2025).
\newblock ``The State of Application Security 2025: How AI Coding Copilots Impact Security Posture.''
\newblock Apiiro.

\bibitem[{Arup/CNN}(2024)]{arup2024}
CNN Business. (2024).
\newblock ``Finance worker pays out \$25 million after video call with deepfake CFO.''
\newblock Available: \url{https://www.cnn.com/2024/02/04/asia/deepfake-cfo-scam-hong-kong-intl-hnk/index.html}

\bibitem[{Boland and Black}(2012)]{juliet2012}
Boland, T. and Black, P. (2012).
\newblock ``Juliet 1.1 C/C++ and Java Test Suite.''
\newblock \emph{IEEE Computer}, 45(10):88--90.

\bibitem[{Carlini et~al.}(2021)]{carlini2021}
Carlini, N., Tramer, F., Wallace, E., Jagielski, M., Herbert-Voss, A., Lee, K., Roberts, A., Brown, T., Song, D., Erlingsson, U., Oprea, A., and Raffel, C. (2021).
\newblock ``Extracting Training Data from Large Language Models.''
\newblock \emph{USENIX Security Symposium}.

\bibitem[{CWE-SANS}(2019)]{cwe2019}
MITRE Corporation and SANS Institute. (2019).
\newblock ``CWE-SANS Top 25 Most Dangerous Software Weaknesses.''
\newblock Available: \url{https://cwe.mitre.org/top25/}

\bibitem[{Dettmers et~al.}(2023)]{dettmers2023}
Dettmers, T., Pagnoni, A., Holtzman, A., and Zettlemoyer, L. (2023).
\newblock ``QLoRA: Efficient Finetuning of Quantized Language Models.''
\newblock \emph{NeurIPS 2023}.

\bibitem[{Grand View Research}(2024)]{grandview2024}
Grand View Research. (2024).
\newblock ``Artificial Intelligence Market Size, Share \& Trends Analysis Report.''
\newblock Grand View Research.

\bibitem[{Greshake et~al.}(2023)]{greshake2023}
Greshake, K., Abdelnabi, S., Mishra, S., Endres, C., Holz, T., and Fritz, M. (2023).
\newblock ``Not What You've Signed Up For: Compromising Real-World LLM-Integrated Applications with Indirect Prompt Injection.''
\newblock \emph{arXiv preprint arXiv:2302.12173}.

\bibitem[{NIST}()]{sard}
NIST.
\newblock ``Software Assurance Reference Dataset (SARD).''
\newblock Available: \url{https://samate.nist.gov/SARD/}

\bibitem[{OWASP}(2025)]{owasp2025llm}
OWASP Foundation. (2025).
\newblock ``OWASP Top 10 for Large Language Model Applications 2025.''
\newblock Available: \url{https://owasp.org/www-project-top-10-for-large-language-model-applications/}

\bibitem[{Perez and Ribeiro}(2022)]{perez2022}
Perez, F. and Ribeiro, I. (2022).
\newblock ``Ignore This Title and HackAPrompt: Exposing Systemic Weaknesses of LLMs through a Global Scale Prompt Hacking Competition.''
\newblock \emph{arXiv preprint arXiv:2311.16119}.

\bibitem[{Russell et~al.}(2018)]{vdisc}
Russell, R., Kim, L., Hamilton, L., Lazovich, T., Harer, J., Ozdemir, O., Ellingwood, P., and McConley, M. (2018).
\newblock ``Automated Vulnerability Detection in Source Code Using Deep Representation Learning.''
\newblock \emph{arXiv preprint arXiv:1807.04320}.

\bibitem[{Veracode}(2025)]{veracode2025}
Veracode. (2025).
\newblock ``2025 GenAI Code Security Report: Assessing the Security of Using LLMs for Coding.''
\newblock Veracode Research.

\bibitem[{Zou et~al.}(2023)]{zou2023}
Zou, A., Wang, Z., Kolter, J.Z., and Fredrikson, M. (2023).
\newblock ``Universal and Transferable Adversarial Attacks on Aligned Language Models.''
\newblock \emph{arXiv preprint arXiv:2307.15043}.

\end{thebibliography}

% ===== Appendix =====
\appendix

\section{Unified Schema}
\label{app:schema}

Both web and AI/ML examples share the unified conversation format. Domain-specific metadata is preserved within each example.

\textbf{Web security example (simplified):}
\begin{lstlisting}[basicstyle=\ttfamily\footnotesize]
{
  "category": "A01-broken-access-control",
  "subcategory": "Insecure Direct Object Reference",
  "severity": "CRITICAL",
  "language": "python",
  "cve": "CVE-2023-XXXXX",
  "conversations": [
    {"role": "human", "content": "..."},
    {"role": "assistant", "content": "..."},
    {"role": "human", "content": "..."},
    {"role": "assistant", "content": "..."}
  ]
}
\end{lstlisting}

\textbf{AI/ML security example (simplified):}
\begin{lstlisting}[basicstyle=\ttfamily\footnotesize]
{
  "id": "llm01-rag-injection-via-llamaindex",
  "metadata": {
    "category": "OWASP LLM Top 10 2025 - LLM01",
    "subcategory": "Indirect Injection",
    "technique": "RAG Document Injection",
    "severity": "CRITICAL",
    "cwe": "CWE-74",
    "lang": "python",
    "owasp_llm_2025": "LLM01"
  },
  "conversations": [
    {"role": "human", "content": "..."},
    {"role": "assistant", "content": "..."},
    {"role": "human", "content": "..."},
    {"role": "assistant", "content": "..."}
  ],
  "quality_score": 94,
  "security_assertions": ["..."],
  "references": [
    {"type": "cve", "id_or_url": "CVE-2024-XXXXX"}
  ]
}
\end{lstlisting}

\section{CWE Coverage (AI/ML Component)}
\label{app:cwe}

The AI/ML component maps to 92 unique CWE identifiers. The 10 most frequent:

\begin{table}[h]
\centering
\begin{tabular}{@{}llc@{}}
\toprule
\textbf{CWE} & \textbf{Description} & \textbf{Files} \\
\midrule
CWE-200 & Information Exposure & 103 \\
CWE-345 & Insufficient Verification of Data Authenticity & 57 \\
CWE-74 & Injection & 42 \\
CWE-94 & Code Injection & 42 \\
CWE-400 & Uncontrolled Resource Consumption & 39 \\
CWE-506 & Embedded Malicious Code & 29 \\
CWE-269 & Improper Privilege Management & 27 \\
CWE-502 & Deserialization of Untrusted Data & 22 \\
CWE-79 & Cross-site Scripting (XSS) & 22 \\
CWE-770 & Allocation of Resources Without Limits & 19 \\
\bottomrule
\end{tabular}
\end{table}

\section{Framework Coverage (AI/ML Component)}
\label{app:frameworks}

Approximate framework usage across the 750 AI/ML examples (many examples use multiple frameworks):

\begin{table}[h]
\centering
\footnotesize
\begin{tabular}{@{}lr|lr|lr@{}}
\toprule
\textbf{Framework} & \textbf{$\sim$n} & \textbf{Framework} & \textbf{$\sim$n} & \textbf{Framework} & \textbf{$\sim$n} \\
\midrule
OpenAI & 514 & Pinecone & 45 & Ollama & 11 \\
Next.js & 230 & ChromaDB & 40 & W\&B & 11 \\
Flask & 179 & Cohere & 40 & Gradio & 10 \\
Anthropic & 175 & Mistral & 36 & MLflow & 10 \\
Dify & 168 & LlamaIndex & 28 & AutoGen & 10 \\
LangChain & 162 & Modal & 27 & Spring Boot & 9 \\
HuggingFace & 130 & Electron & 24 & gRPC & 9 \\
Express & 102 & Weaviate & 21 & AWS Bedrock & 8 \\
Together AI & 90 & CrewAI & 14 & Streamlit & 7 \\
FastAPI & 83 & DeepSeek & 14 & vLLM & 7 \\
React & 59 & Vercel AI SDK & 14 & +14 more & $<$5 ea. \\
\bottomrule
\end{tabular}
\end{table}

\end{document}